\documentclass[reprint]{revtex4-2}

\usepackage{amsmath,amssymb,amsfonts}
\usepackage{graphicx}
\usepackage[hidelinks]{hyperref}

\newcommand{\bx}{{\bf x}}
\newcommand{\X}{{\bf X}}
\newcommand{\E}{\mathbb{E}}
\newcommand{\calL}{\mathcal{L}}
\newcommand{\calS}{\mathcal{S}^\tau}
\newcommand{\shortlag}{\rho}
\newcommand{\T}{\sigma}

\begin{document}

\title{Machine learning kinetics from molecular dynamics data}
\author{Jonathan Weare}
\email{weare@nyu.edu}
\affiliation{Courant Institute School of Mathematics, Computing, and Data Science,\\ New York University, New York, New York 10012, United States}

\author{Aaron R. Dinner}
\email{dinner@uchicago.edu}
\affiliation{Department of Chemistry and James Franck Institute, University of Chicago, Chicago, Illinois 60637, United States}

\begin{abstract}
Most molecular transitions occur on time scales far beyond direct molecular dynamics simulations. The committor, the probability that a configuration reaches a product state before a reactant state, is a central kinetic statistic, providing a mechanism-independent reaction coordinate and a foundation for transition path theory and the calculation of rates. This review surveys modern approaches for estimating the committor and related kinetic statistics from molecular simulations, with an emphasis on self-supervised methods that learn solutions of their defining dynamical equations rather than relying on labeled shooting data. We develop a common operator viewpoint connecting generator-based partial differential equations, variational principles, Markov state models, dynamical Galerkin approximation, and neural networks. Empirical and theoretical evidence points to the efficiency of these methods. We provide theoretical and practical guidance for realizing their full potential in applications, including strategies for treating non-Markovian effects and for sampling. We conclude by identifying opportunities for further research, including connections to reinforcement learning and generative modeling.
\end{abstract}

\maketitle

\section{Introduction}

Molecular dynamics (MD) simulations are essential for connecting molecular structure to dynamics and function measured experimentally \cite{karplus2002molecular,hollingsworth2018molecular}.  Not only can simulations fully resolve microscopic dynamics, but energies and forces can be computed, and force field parameters can be varied systematically to probe their roles. This information can reveal mechanisms and guide molecular design.  However, atomic-resolution simulations remain limited to time scales of $\sim$10$^{-5}$--10$^{-4}$ s \cite{lindorfflarsen2011how,jensen2012mechanism}, which prevents them from directly accessing the time scales of experiments and the  phenomena that they probe ($\sim$10$^{-3}$--10$^4$ s for the molecular examples that we describe in this review). This time scale gap makes observing one event in unbiased simulations prohibitively computationally costly, let alone the many needed for statistics and computing experimental observables.
% or "rare-event problem"
% specific examples: protein folding, allosteric transitions, ligand binding, ion-channel gating

In many cases, the time scale gap results from systems spending long times (many MD steps) in metastable states (reactants, products, and intermediates) interrupted by comparatively brief transitions between those states \cite{fleming1990chemical,min2005fluctuating,schuler2008protein,lindorfflarsen2011how}. This observation is the basis for almost all methods for enhanced sampling of events:  simulations are biased toward key states with low probabilities either directly through progress variables or indirectly (e.g., by increasing temperature) and then the results are reweighted to recover unbiased statistics.  While there are now well-established enhanced sampling methods for not only equilibrium averages \cite{allen1987computer,frenkel2002understanding,henin2022enhanced} but also dynamical averages \cite{zuckerman2017weighted,bolhuis2002transition,allen2009forward,dickson2010enhanced}, identifying suitable progress variables can be challenging for complex systems \cite{ma2005automatic,li2014recent,bittracher2018data,brandt2018machine,gkeka2020machine,bussi2020using,jung2023machine,ma2025reaction}.  

This review discusses methods for computing statistics that can be used to identify suitable progress variables and to characterize states and pathways contributing to transitions.  We focus particularly on \emph{the committor} $q(\bx)$---the probability that a trajectory starting from microscopic state $\bx$ reaches the product state ($B$) before the reactant state ($A$) under the unbiased dynamics---and how it can be estimated efficiently from MD simulation data using machine learning methods.  By definition $q(\bx) = 0$ for $\bx\in A$ and $q(\bx) = 1$ for $\bx\in B$; $q=1/2$ (equal likelihood of next reaching $A$ and $B$) provides a definition of transition states that requires no a priori knowledge of a mechanism \cite{du1998transition,bolhuis2000reaction,ma2005automatic}.  The committor is arguably the optimal reaction coordinate
\cite{du1998transition,hummer2004from,berezhkovskii2013diffusion,li2014recent,banushkina2016optimal,peters2016reaction,elber2017calculating,berezhkovskii2019committors,ma2025reaction}, and it is the central quantity in transition path theory for computing reaction rates \cite{e2005transition,e2006towards,vandeneijnden2006transition,metzner2009transition,e2010transition}, as we discuss below.

% We need to establish "rules" for italicization

\subsection{Rate theories}
\label{sec:tsttpt}
The traditional starting point of chemical reaction kinetics is \emph{transition state theory} (TST) \cite{chandler1978statistical,laidler1983development,fleming1990chemical,zhou2010rate,peters2017reaction}.  In TST, the rate constant $k$ is proportional to the equilibrium flux crossing a hypersurface that separates the reactant and product states.  TST is exact when there are no hypersurface recrossings, but this is rarely the case for reactions in condensed phases, so TST generally overestimates rates. Kramers'  and subsequent theories improve on TST by accounting for the friction and curvature of the barrier
\cite{hanggi1990reaction,berezhkovskii2005one,zhou2010rate,peters2017reaction,acharya2021rate,hong2022introduction}, but they still assume that the transition is dominated by a single, well-defined saddle point of the (free) energy.  This assumption is violated in many systems studied today, ranging from surface-catalyzed reactions to protein conformational changes \cite{mccandler2026markov,antoszewski2021kinetics,jeong2024analysis,guo2024dynamics}. These transitions can involve many competing pathways with broad and diffuse transition-state regions.

\emph{Transition path theory} (TPT) \cite{e2005transition,e2006towards,vandeneijnden2006transition,metzner2009transition,e2010transition} is a mechanism-agnostic framework for kinetics.  TPT characterizes the statistics of the ensemble of reactive trajectories (or equivalently transition paths).  The information about reactive trajectories is encoded in the stationary distribution, the (forward) committor, and the backward committor---the probability that, looking backward in time from $\bx$, the trajectory last came from $A$ rather than $B$---all of which are functions of microscopic states, not paths. As we discuss below, for reversible dynamics, the unimolecular rate $r_{AB}$, the number of transitions from $A$ to $B$ per unit time, can be obtained by minimizing a variational expression for trial committor functions (see Section~\ref{sec:DF}). Note that $r_{AB}$ is symmetric in $A$ and $B$ ($r_{AB}=r_{BA}$) and should be distinguished from the TPT rate $k_{AB}$, which is obtained from $r_{AB}$ by dividing by the proportion of time the process spends having last come from $A$, i.e., by the stationary average value of the backward committor. Henceforth, ``rate'' in this review refers to $r_{AB}$.

For general dynamics (including dynamics that violate detailed balance), TPT computes the rate from an integral over a dividing hypersurface of the reactive flux expressed in terms of the stationary distribution and the forward and backward committors; in contrast to the rate theories described above, the dividing hypersurface is arbitrary, and the underlying trajectories may have any number of recrossings. The level sets of an approximate committor are natural choices in practice, and one can average over them to further suppress noise in rate estimates~\cite{strahan2021long}. Some loss functions for training the committor covered in this review can also be used to estimate $r_{AB}$ (see Sections \ref{sec:DF} and \ref{sec:VCN}). Ref.\ \citenum{berezhkovskii2019committors} connects TPT to Kramers' theory.

\subsection{Roadmap}
\label{sec:roadmap}

TPT is a framework for computing reactive fluxes and rates given the stationary distribution and committors, but it does not provide a prescription for estimating these quantities from MD simulation data.  As noted above, methods for computing the stationary distribution are well-established, at least in reversible systems \cite{allen1987computer,frenkel2002understanding,henin2022enhanced} (see \cite{dickson2010enhanced} for irreversible systems).  We thus focus on the committor here.

Initially $q(\bx)$ was computed by shooting:  for a configuration $\bx$, one launches $M$ independent trajectories with momenta drawn from the Maxwell--Boltzmann distribution and counts the number that reach $B$ before $A$ \cite{du1998transition,bolhuis2002transition,peters2016reaction}.   This approach is computationally costly because trajectories must be run all the way to $A$ or $B$ and accurate estimates require many trajectories for each $\bx$ tested. Specifically, because the outcomes are binary ($A$ or $B$), the error follows the formula for the Bernoulli distribution, $\sqrt{q(1-q)/M}$ \cite{peters2006using,peters2016reaction} (though the error can be reduced with empirical Bayes \cite{gurumoorthi2026improving}). The error not only scales as $1/\sqrt{M}$ but peaks at the transition state $q=1/2$, which dominates rates and is of greatest mechanistic interest.
% [ref] added: shooting for the committor

Ma and Dinner \cite{ma2005automatic,hu2008two} showed that one can regress on such data, which both combines information from multiple $\bx$ and yields models of the committor in terms of candidate physical variables.  Peters and Trout \cite{peters2006obtaining,peters2007extensions} improved the data efficiency of this procedure through the introduction of aimless shooting, and they instead evaluated models with likelihood maximization. Jung {\it et al.}\ \cite{jung2023machine} made the sampling adaptive, alternating between training a model of the committor on shooting outcomes and selecting new shooting points that the current model places near the transition state.
As we discuss in Section \ref{sec:shooting}, these approaches have been used to learn the committor for a wide range of transitions.
An advantage of methods that regress on small numbers of physical variables is that the resulting models are interpretable, but  their accuracy is ultimately limited by the library of candidate physical variables. The requirement that the trajectories reach the reactant or product state can be hard to meet when systems have long-lived intermediates and/or highly diffusive dynamics.

These issues motivate approaches in which one represents the committor by a general parameterized functional form and then determines the values of the parameters so that the resulting function satisfies a defining equation of the committor.  In the case that the dynamics are fully specified, the defining equation can take the form of a partial differential equation (PDE) \cite{gardiner2009stochastic,pavliotis2014stochastic,peters2017reaction}.  For example, for overdamped Langevin dynamics, the committor is the stationary solution to the backward Kolmogorov equation (and the associated operator is the adjoint to the forward Kolmogorov/Fokker--Planck operator). When the PDE is high-dimensional, as is generally the case, it cannot be treated by traditional (e.g., grid-based) numerical approaches, whose computational costs scale exponentially with the number of variables. A successful alternative is to represent the committor by a neural network (NN), as we discuss in Section \ref{sec:PDE}.  

However, when the analysis focuses on a reduced set of variables (e.g., by neglecting those of the solvent) for interpretability and/or statistical power, the partial differential form is not available, and the operator must instead be formulated as an expectation over the dynamics. The transition matrix in Markov state models \cite{bowman2013introduction,swope2004describinga,prinz2011markov,husic2018markov,wang2018constructing}, which is computed from statistics of state-to-state transitions within a lag time $\tau$ in molecular dynamics simulations, can be viewed as an expansion of this operator in a basis of indicator functions. Dynamical Galerkin approximation and related methods \cite{thiede2019galerkin,strahan2021long,khoo2019solving,cao2026continuous} generalize the expansion of the operator and committor to allow for other basis functions and to account for the boundary conditions $q(\bx) = 0$ for $\bx\in A$ and $q(\bx) = 1$ for $\bx\in B$. These methods can achieve more accurate committors than MSMs, but they require choosing basis functions and the features on which they depend. NNs that learn functions that satisfy the operator equation and/or basis functions offer routes to going beyond manual feature engineering and are an active area of research \cite{strahan2023inexact}, as we discuss in Section \ref{sec:trajectory}.

To understand the distinctions between methods, it can be useful to categorize them from a machine learning perspective. The regression and likelihood maximization approaches described above are supervised in the sense that the tested configurations are paired with outcomes of independent MD simulations. The remaining approaches in this section can be considered self-supervised in that they test for satisfaction of equations and boundary conditions without requiring independent MD simulations; these approaches are the main focus of this review. Neural-network approaches that use fixed-point iteration to solve operator equations for finite lag times \cite{strahan2023inexact} are also closely related to temporal difference approaches in reinforcement learning  \cite{sutton2018reinforcement}. 

In the remainder of this review, we introduce the mathematical framework for the self-supervised methods, which rest on the Markov approximation. Using this framework, we describe the methods more precisely and give examples of applications. We then discuss ways to go beyond the Markov approximation.  We argue that good coverage in the transition region is important and discuss how it can be achieved through adaptive sampling.  We end by briefly discussing alternative approaches and identifying areas that can be fruitful for further research.

\section{PDE approximation of the committor}
\label{sec:PDE}

Our starting point for computing the committor is its definition:  $q(\bx)$ is the probability that a system in state $\bx$ reaches $B$ before $A$.  This probability can be obtained from the flux into $B$ from trajectories that start in $\bx$. An equation for this flux can be obtained from the backward Kolmogorov equation for the conditional probability $p({\bf y},t\mid \bx)$ \cite{gardiner2009stochastic,peters2017reaction}:
% [ref] petersbook; Gardiner's Handbook of Stochastic Methods is NOT in master.bib
% which is commonly used for terminal value (and thus first passage) problems.
\begin{equation}\label{eq:backward}
    \frac{\partial }{\partial t} p({\bf y},t\mid \bx) = \calL p({\bf y},t\mid \bx).
\end{equation}
In this equation, $\bx$ acts as the argument to the function $p({\bf y},t\mid \bx)$, and  the state $\bf y$ acts as a parameter.  That is, the operator $\calL$ acts on functions of $\bx$, $f(\bx)$.
%That is, the backward equation describes the 

The operator $\calL$ is the infinitesimal generator of the dynamics; it plays a role in stochastic dynamics that is analogous to the Liouville operator in classical mechanics (in fact, $\calL$ contains the Liouville operator when there is a deterministic component to the dynamics) \cite{pavliotis2014stochastic,zwanzig2001nonequilibrium}.  Specifically for an overdamped Langevin dynamics with force derived from a potential $V(\bx)$, stationary probability $\pi(\bx)\propto e^{-\beta V(\bx)}$, position-dependent diffusion tensor $D(\bx)$, and inverse temperature $\beta$,
\begin{equation}\label{eq:overdamped}
    \mathcal{L}f(\bx) = e^{\beta V(\bx)}\,\nabla\cdot\Big[e^{-\beta V(\bx)}\,D(\bx)\,\nabla f(\bx)\Big].
\end{equation}
More generally, for the Markov process $\X_t$ at time $t$,
\begin{equation}\label{eq:generator}
\calL f(\bx) = \lim_{\tau\to 0^+}\frac{\E[f(\X_\tau)\mid \X_0 = \bx] - f(\bx)}{\tau},
\end{equation}
where $\E$ denotes an expectation.  Eq.\ \eqref{eq:generator} is essentially a definition of the time derivative, consistent with \eqref{eq:backward}.

Integrating the flux equation, which has a form that parallels \eqref{eq:backward} with the flux replacing $p$, gives \cite{peters2017reaction,onsager1938initial,tachiya1978general,schulten1981dynamics}
\begin{equation}\label{eq:Lq=0}
    \calL q(\bx) = 0 \quad \text{for} \quad \bx \notin (A\cup B),
\end{equation}
with $q(\bx) = 0$ for $\bx\in A$ and $q(\bx) = 1$ for $\bx\in B$.
Physically, \eqref{eq:Lq=0} says that, an infinitesimal time later along each trajectory from $\bx$, the expected end state of the trajectory ($A$ or $B$) is conserved.  

% Added 2026-08-03
Other kinetic statistics can be written in terms of equations of the generator as well.  For example, the mean first passage time (MFPT) to $B$ from microscopic state $\bx$ satisfies \cite{pavliotis2014stochastic,peters2017reaction}
\begin{equation}\label{eq:Lmfpt}
    \calL m(\bx) = -1 \quad \text{for} \quad \bx \notin B
\end{equation}
with $m(\bx)=0$ for $\bx\in B$.  That is, an infinitesimal time later along each trajectory, the MFPT is one time unit less.  More generally, this approach can be extended to conditional expectations of the form \cite{pavliotis2014stochastic}
\begin{equation}\label{eq:general_bvp}
  u(\bx)
  = \E\!\left[
      \Psi(\X_{\T}) + \int_0^{\T} R(\X_t)\,dt \,\Big|\, \X_0 = \bx
    \right],
\end{equation}
where $\T = \min\{t : \X_t\in \Omega\}$ is the time of first reaching $\Omega$, and $R(\bx)$ and $\Psi(\bx)$ are running and terminal ``reward'' functions. 
The committor corresponds to the specific choices $\Omega = A\cup B$, $R(\bx) = 0$,
and $\Psi(\bx) = 1$ for $\bx\in B$ and $0$ elsewhere; the MFPT corresponds to  $\Omega = B$, $R(\bx) = 1$, and $\Psi(\bx) = 0$, reflecting that no reward is collected upon reaching $B$, but each unit of time spent before arrival contributes one unit to the total.

\subsection{Physics-informed NNs (PINNs)}
\label{sec:PINN}

When an explicit form for $\calL$ such as \eqref{eq:overdamped} is available, \eqref{eq:Lq=0} is a PDE.  In this case, one can leverage methods for solving high-dimensional PDEs \cite{sirignano2018dgm,han2018solving,e2018deep,hermann2020deep,han2020solving}.
% [ref] refs -> high-dimensional PDE solvers
The general strategy is to represent trial solutions to  PDEs  by parameterized functions ($q_\theta$ with parameters $\theta$ for the committor) and to adjust the parameters to minimize the PDE residuals ($\calL q_\theta$ for the committor) \cite{karniadakis2021physics,raissi2019physics}.  Software libraries now make NNs convenient choices for the parameterized functions (though others are possible \cite{lai2018point,chen2023committor,evans2022computing,lucke2022tgedmd}).  In the case of the committor, the part of the loss that enforces the PDE and thus makes the training \emph{physics-informed} (PI) takes the form \cite{strahan2023predicting} % eq 22 in two-trajectory paper is PINN loss
\begin{equation}\label{eq:PINN_loss}
  J^{\mathrm{PINN}}
  = \E_{\X_0\sim\mu}\!\big[(\calL q_\theta(\X_0))^2\big],
\end{equation}
where the subscript on the expectation indicates the distribution over which it is computed; here, $\mu(\bx)$ is a user-chosen distribution with support on $\bx\notin(A\cup B)$. In practice, $\mu(\bx)$ is represented by  $M$ independent samples $\X_0^{(1)},\ldots,\X_0^{(M)}\notin(A\cup B)$. 

In addition, it is necessary to enforce the boundary conditions. A simple choice is to add to the loss above a term of the form \cite{strahan2023predicting,li2022semigroup,rotskoff2022active,khoo2019solving,chen2023committor}
\begin{equation}
    J^{\mathrm{BC}}=\E_{\X_0\sim A}[(q_\theta(\X_0))^2] + \E_{\X_0\sim B}[(1-q_\theta(\X_0))^2].
\end{equation}
The relative weights of $J^{\mathrm{PINN}}$ and $J^{\mathrm{BC}}$ can be tuned by multiplying one term by a hyperparameter.
Alternatively, one can use a construction that imposes the boundary conditions by masking the output of the neural network within the metastable states, e.g., $q_\theta(\bx) = (1-\chi_A(\bx))\,[(1-\chi_B(\bx))\,\tilde{q}_\theta(\bx)
+ \chi_B(\bx)]$, where the tilde indicates an intermediate output and $\chi_A$ is an indicator function that equals $1$ inside $A$ and $0$ otherwise \cite{li2019computing,chen2023discovering}; in some cases \cite{li2019computing}, $\chi_A$ is smoothed. 
With masking, $q_\theta(\bx)$ satisfies the boundary conditions exactly for any $\theta$ but can exhibit discontinuities at the boundaries of $A$ and $B$; 
adding boundary penalty terms can mitigate these discontinuities \cite{lorpaiboon2026exact,megias2025iterative,chen2026following}.

The PINN loss is conceptually important, and it has the advantage that the distribution $\mu(\bx)$ can be chosen freely. However, the second-order derivatives required (see \eqref{eq:overdamped}) can be problematic numerically:  errors tend to be amplified, NN activations must be smooth, and the computational cost can scale quadratically with system dimension.  For these reasons, researchers generally use other methods that we describe in the remainder of this review.

\subsection{Dirichlet form}
\label{sec:DF}

When the dynamics are microscopically reversible (i.e., satisfy detailed balance with respect to the stationary distribution $\pi$), the generator is self-adjoint \cite{pavliotis2014stochastic}:
\begin{equation}
  \E_{\X_0\sim\pi}[f(\X_0)\,\calL g(\X_0)]
  = \E_{\X_0\sim\pi}[g(\X_0)\,\calL f(\X_0)]
  \label{eq:self_adjoint}
\end{equation}
for sufficiently smooth functions $f$ and $g$. In this case, the committor can be obtained from a variational principle: among trial functions $q_\theta$ satisfying $q_\theta(\bx) = 0$ for $\bx\in A$ and $q_\theta(\bx) = 1$ for $\bx\in B$, the exact committor minimizes the \emph{Dirichlet form} \cite{pavliotis2014stochastic}
\begin{equation}\label{eq:DF_loss}
  J^{\mathrm{DF}}[q_\theta]
  = -\E_{\X_0\sim\pi}[q_\theta(\X_0)\,\calL q_\theta(\X_0)].
\end{equation}
Functionally differentiating \eqref{eq:DF_loss} and integrating by parts shows that the minimizer satisfies \eqref{eq:Lq=0}.

Two further properties of the Dirichlet form hold regardless of whether the dynamics are reversible. First, evaluated at the exact committor, it equals the steady-state rate of transitions from $A$ to $B$ \cite{vandeneijnden2006transition,e2006towards}:
\begin{equation}
  r_{AB} = -\E_{\X_0\sim\pi}[q(\X_0)\calL q (\X_0)].
  \label{eq:rate_DF}
\end{equation}
This result can be obtained by writing the rate in terms of the reactive flux through a dividing hypersurface. Second, for any generator of the form $\mathcal{L}=a\cdot \nabla  + D:\nabla\nabla$ with stationary distribution $\pi$, where $a$ is the drift and $\nabla\nabla$ is the matrix of second derivatives (e.g., \eqref{eq:overdamped}), the Dirichlet form can be rewritten as \cite{pavliotis2014stochastic}
\begin{equation}\label{eq:DF_gradient}
  J^{\mathrm{DF}}[q_\theta]
  = \E_{\X_0\sim\pi}\!\big[
      \nabla q_\theta(\X_0)^\top D(\X_0)\,\nabla q_\theta(\X_0)
    \big].
\end{equation}
%The overdamped generator \eqref{eq:overdamped} has this form.
Combining these results with the variational principle, $J^{\mathrm{DF}}[q_\theta]\geq J^{\mathrm{DF}}[q] = r_{AB}$ for reversible dynamics, so the Dirichlet form overestimates the rate for any $q_\theta(\bx)\neq q(\bx)$. This inequality can be interpreted as saying that the current associated with the exact committor, $\pi(\bx)D(\bx)\nabla q(\bx)$, carries reactive probability from $A$ to $B$ with minimum possible dissipation.

The form in \eqref{eq:DF_gradient} is widely used (typically with $D$ assumed to be position-independent, so it can be factored out of the expectation) \cite{khoo2019solving,li2019computing,lai2018point,rotskoff2022active,hasyim2022supervised,chen2023committor,singh2023variational,yuan2024optimal}.  Applications include protein folding \cite{kang2024computing,trizio2025everything}, chemical reactions \cite{kang2024computing,trizio2026ceci}, ion association/dissociation \cite{kang2026committors}, crystal nucleation \cite{deng2026role,trizio2026ceci}, host-guest and protein-ligand binding \cite{trizio2025everything,kang2026committors,trizio2026ceci}, and membrane fusion \cite{rossi2026let}.
Compared with the PINN loss \eqref{eq:PINN_loss}, \eqref{eq:DF_gradient} requires only first-order derivatives of $q_\theta(\bx)$ rather than second-order derivatives. However, because  \eqref{eq:DF_gradient} ultimately derives from \eqref{eq:self_adjoint}, it requires that data be drawn from (or reweighted to) the stationary distribution $\pi(\bx)$.  This requirement is statistically undesirable because it puts low weight on transition states, as we discuss further in Section \ref{sec:sampling}.
Furthermore, minimizing the Dirichlet form to learn the committor is restricted to reversible dynamics; for nonequilibrium systems, the PINN loss or an appropriate trajectory-based method (Section~\ref{sec:trajectory}) must be used.

A practical limitation shared by \eqref{eq:PINN_loss}, \eqref{eq:DF_loss}, and  \eqref{eq:DF_gradient} is the requirement to evaluate the generator $\calL$ explicitly. When one wishes to model the committor as a function of a reduced set of variables  $\boldsymbol{\xi}(\bx)\in\mathbb{R}^m$ with $m$ much small than the full dimension of the system (e.g., pairwise distances between selected atoms), the projected dynamics are governed by a generalized Langevin equation, which is non-Markovian: the dynamics of $\boldsymbol{\xi}$ at time $t$ depend on its history, not just its instantaneous value. Even when the memory is sufficiently short that the dynamics can be approximated by an ordinary Langevin equation, the diffusion tensor $D(\bx)$ is not directly available from the force field. Studies that apply \eqref{eq:PINN_loss}, \eqref{eq:DF_loss}, or \eqref{eq:DF_gradient} to selected atoms implicitly assume $D$ is constant and isotropic, which is generally not the case. While various approaches for estimating $D(\bx)$ from data exist \cite{straub1987calculation,straub1990spatial,crouzy1994molecular,im2002ions,hummer2005position,ma2006dynamic,crommelin2011diffusion,ghysels2017position,harris2026deep}, 
they introduce statistical errors that propagate to the committor and the TPT statistics computed from it. This issue motivates the methods in the next section, which instead estimate kinetic statistics from trajectory data. Throughout the review we write the committor and other statistics as functions of the full microscopic state, $q(\bx)$, even when the discussion concerns a reduced set of variables $\boldsymbol{\xi}(\bx)$, to keep the notation uniform across sections.

\section{Learning the committor from trajectory data}
\label{sec:trajectory}

There are many cases where it is not possible, or is undesirable, to evaluate the generator $\calL$ explicitly.  These include not only the common case of analysis based on a reduced set of variables discussed above but also analysis of data from black-box simulators, experiments, and observations.  In this section, we discuss approaches to compute the committor and other kinetic statistics from trajectory data saved at finite intervals $\tau$, without an explicit model for the dynamics.

\subsection{Markov state models}
\label{sec:MSM_DGA}

\emph{Markov state models} (MSMs) \cite{bowman2013introduction,swope2004describinga,prinz2011markov,husic2018markov,wang2018constructing} are among the oldest and most widely used methods for approximating the long-time kinetics of
molecular systems from trajectory data. One discretizes the space of dynamical variables into a finite number of states $\{\Omega_1,\ldots,\Omega_N\}$ and estimates the transition probability matrix elements $T_{ij}^\tau = P[\X_\tau\in \Omega_j\mid \X_0\in \Omega_i]$ from the data. Statistics such as stationary distributions, relaxation time scales, and MFPTs can be computed from equations of the matrix ${\bf T}^\tau$. In particular, the committor can be approximated by solving the linear system
\begin{equation}\label{eq:MSMq}
  %\sum_j (T_{ij}^\tau - \delta_{ij})\,q_j = 0 \hspace{0.75em} \text{for} \hspace{0.75em} i\hspace{0.75em}\text{with}\hspace{0.75em}  \Omega_i\notin A\cup B,
  \sum_j (T_{ij}^\tau - \delta_{ij})\,q_j = 0 \quad \text{for} \quad \{i: \Omega_i\notin A\cup B\},
\end{equation}
where $\delta_{ij}$ is the Kronecker delta, and $q_i = 0$ for $\Omega_i\in A$ and $q_i = 1$ for $\Omega_i\in B$.
Applications of this approach include peptide folding \cite{noe2009constructing,banerjee2014transition}, dimerization \cite{qiao2016dynamics}, functional protein conformational changes \cite{malmstrom2014application,meng2016transition,dandekar2023markov}, ice nucleation \cite{li2021temperature}, and virus capsid assembly \cite{trubiano2024markov}.

One can view ${\bf T}^\tau$ as the instantiation of the transition operator 
\begin{equation}\label{eq:Toperator}
    {\cal T}^\tau f(\bx) = \E[f(\X_\tau)\mid \X_0 = \bx]
\end{equation}
in a basis of indicator functions on the states $\{\Omega_1,\ldots,\Omega_N\}$ \cite{bittracher2018data,thiede2019galerkin}. From this perspective, \eqref{eq:MSMq} is the finite-$\tau$ analog of \eqref{eq:Lq=0} with the definition of $\calL$ in \eqref{eq:generator}.  As in the PINN approach above, an advantage of MSMs is that one has the freedom to choose the sample distribution $\mu(\bx)$ because the statistics are conditioned on the initial state.  In practice, the choice of $\mu(\bx)$ affects the results in ways that depend on the choice of states and the approach used to estimate $T_{ij}^\tau$ from finite data (both of which remain active areas of research) \cite{chodera2007automatic,deuflhard2005robust,bowman2009progress,trendelkampschroer2015estimation,husic2018markov,arbon2024markov,aristoff2026markov}.

However, even in the limit of infinite data, MSM estimates for the committor have two sources of error. The first is discretization:  every $\bx$ within state $\Omega_i$ is assigned committor value $q_i$.  This source of error can be mitigated by using a fine discretization where the committor changes rapidly (i.e., the transition state region), which in practice, with finite data, spreads the data over more states.  The second source of error holds even for an infinitely fine discretization:  there is a $\tau$-dependent bias from the
finite-$\tau$ approximation of \eqref{eq:Lq=0} \cite{strahan2021long,tuchkov2025error,lorpaiboon2026exact}. 
In the next section, we address this finite-$\tau$ error by introducing an operator that defines exact statistics; we then discuss various approaches for solving equations of this operator with expressive continuous representations, which address the discretization error.

\subsection{The stopped transition operator}

The finite-$\tau$ error discussed above arises when trajectories that pass through the metastable states $A$ and $B$ contribute to $T_{ij}^\tau$.  This issue can be addressed by defining an analog of \eqref{eq:Toperator} where the trajectories are \emph{stopped} when they reach the metastable states \cite{strahan2021long}:
\begin{equation}
  \mathcal{S}^\tau f(\bx)
  = \E[f(\X_{\tau\wedge \T})\mid \X_0 = \bx],
  \label{eq:stopped_operator}
\end{equation}
where $\T = \min\{t: \X_t \in \Omega\}$ is the time of first reaching a terminal state $\Omega$ and $\tau\wedge \T = \min\{\tau,\T\}$; for the committor $\Omega=A\cup B$. 

The committor equation becomes \cite{strahan2021long}
\begin{equation}\label{eq:fixed_point}
  \mathcal{S}^\tau q(\bx) - q(\bx) = 0\quad\text{for}\quad \bx \notin (A\cup B),
\end{equation}
with  $q(\bx) = 0$ for $\bx\in A$ and $q(\bx) = 1$ for $\bx\in B$.
The physical interpretation is the same as that discussed below \eqref{eq:Lq=0}, but now the time along each trajectory is finite ($\tau$ or $\T$, whichever comes first).  In analogy to equations of $\calL$, equations of $\calS$ can be written for other statistics of the form in \eqref{eq:general_bvp} \cite{strahan2021long}. For example, the analog of the MFPT equation \eqref{eq:Lmfpt} is 
\begin{equation}\label{eq:trajmfpt}
    \mathcal{S}^\tau m(\bx) - m(\bx) = -\E[\tau\wedge\T\mid \X_0=\bx]
\end{equation}
with $m(\bx) = 0$ for $\bx\in B$.
%In contrast to equations of $\calL$ (or ${\mathcal T}^\tau$ for problems involving terminal states), equations of $\calS$ are exact for all finite $\tau$, which is important for taking $\tau$ sufficiently long that the dynamics are effectively Markovian.  
In contrast to equations of $\mathcal{T}^\tau$, which, as discussed above, are biased when trajectories are not stopped upon reaching terminal states, equations of $\mathcal{S}^\tau$ are exact for all finite $\tau$.
This feature is important for taking $\tau$ sufficiently long that the dynamics are effectively Markovian (as discussed further in Section \ref{sec:nonmarkov}).  
The methods that we now discuss solve equations of $\calS$ using trajectory data without an explicit form for the operator.

\subsection{Dynamical Galerkin approximation}
\label{sec:DGA}

\emph{Dynamical Galerkin approximation} (DGA), which was first introduced in the context of equations of $\calL$ \cite{thiede2019galerkin}, solves equations of $\calS$ such as \eqref{eq:fixed_point} \cite{strahan2021long} by expanding statistics (here, the committor) in a basis:
% [ref] thiede2019dynamical
%
\begin{equation}
  q_\theta(\bx) = \phi_0(\bx) + \sum_{\ell=1}^L c_\ell\,\phi_\ell(\bx),
  \label{eq:DGA_ansatz}
\end{equation}
where $\phi_0(\bx)$ accounts for the boundary conditions, and $\phi_\ell(\bx)=0$ for $\ell > 0$ and $\bx\in A\cup B$; $c_\ell$ are coefficients.  Substituting \eqref{eq:DGA_ansatz} into \eqref{eq:fixed_point}, multiplying by $\phi_k(\bx)$, and taking the expectation with respect to the sample distribution $\mu(\bx)$ yields the linear system
%
% \begin{equation}
%   ({\bf C}^\tau - {\bf C}^0){\bf c} = \mathbf{b}^\tau,
%   \label{eq:DGA_linear}
% \end{equation}
\begin{equation}
  {\bf C}^\tau{\bf c} = \mathbf{b}^\tau,
  \label{eq:DGA_linear}
\end{equation}
with
\begin{align}
  % C^t_{k\ell} \label{eq:Ctau}
  % &= \E_{\X_0\sim\mu}[\phi_k(\X_0)\,\phi_\ell(\X_{t\wedge \T})]
  %  \quad \text{for} \quad t=0, \tau\\
 C^\tau_{k\ell}
  &= \E_{\X_0\sim\mu}\big[\phi_k(\X_0)\,(\phi_\ell(\X_{\tau\wedge \T})- \phi_\ell(\X_{0}))\big]
  \label{eq:Ctau}\\
  b_k^\tau    \label{eq:bvec}
  &= -\E_{\X_0\sim\mu}[
      \phi_k(\X_0)\,(\phi_0(\X_{\tau\wedge \T})-\phi_0(\X_0))
    ].
\end{align}
For other statistics of the form in \eqref{eq:general_bvp}, $b_k^\tau$ includes an additional term involving the running reward integral \cite{strahan2021long}. %There is not assumption of detailed balance, so DGA can be applied to irrersible dynamics.

The entries of ${\bf C}^\tau$ and $\mathbf{b}^\tau$ can be estimated from a set of trajectory segments $\{(\X_0^{(m)}, \X_{\tau\wedge \T^{(m)}}^{(m)})\}_{m=1}^M$, where each segment starts at an initial condition $\X_0^{(m)}$ drawn from a sample distribution $\mu(\bx)$ and
runs until time $\tau$ or until reaching $A\cup B$, whichever comes first.  As for the PINN loss and MSMs, $\mu(\bx)$ can be user-chosen, and knowledge of its form is not needed; again, this enables freedom to concentrate the sampling where it is needed most.  Once  ${\bf C}^\tau$ and $\mathbf{b}^\tau$ are estimated, the coefficient vector {\bf c} can be determined from \eqref{eq:DGA_linear}, and the statistic can be computed from \eqref{eq:DGA_ansatz}.
In particular, given an approximate committor, the rate can be computed as an average over its level sets, as described in Section \ref{sec:tsttpt} \cite{strahan2021long,jeong2024analysis}, or using the estimators in Section \ref{sec:VCN}.

When the $\phi_\ell(\bx)$ are indicator functions for discrete sets of microscopic states, DGA reduces to MSMs built on the stopped rather than unstopped transition matrix; that is, the matrix is constructed to account for statistic-specific boundary conditions. However, other choices of basis functions can be made to avoid the discretization error discussed above.
Strahan {\it et al.}\ \cite{strahan2021long} introduced a construction in which pairwise distances are multiplied by a function that vanishes on $A$ and $B$; the resulting functions are then grouped into vectors and orthogonalized (whitened) through singular value decomposition (SVD). 
In related approaches, Khoo {\it et al.}\ \cite{khoo2025fast} and Cao and Huang \cite{cao2026continuous} consider polynomial, Fourier, and radial basis functions. Basis functions have also been learned: Thiede {\it et al.}\ \cite{thiede2019galerkin} used diffusion maps, and Strahan {\it et al.}\ \cite{strahan2023inexact} used subspace iteration to train NNs.  We discuss the latter in Section \ref{sec:subspace}.

\subsubsection{Applications}
\label{sec:applications}

Here, we showcase four simulation studies that illustrate how features of DGA enable accurate committor estimates and, in turn, mechanistic insight.

\paragraph{Trp-cage folding.}  
Trp-cage is a designed 20-residue fast-folding protein that has been studied extensively both experimentally and computationally (see Ref.~\citenum{strahan2021long} and references therein). Shaw and co-workers previously generated a 208 $\mu$s trajectory \cite{lindorfflarsen2011how}, but it contains only 12--31 folding/unfolding events (depending on state definitions) \cite{lindorfflarsen2011how,deng2013how,sidky2019high} and thus little sampling of the transition state region. To address this limitation, Strahan {\it et al.}\ \cite{strahan2021long} generated a data set of short (30 ns) trajectories (totaling 30 $\mu$s) distributed uniformly in selected pairs of variables, adding trajectories where replica-exchange umbrella sampling (REUS) revealed the unfolded state was undersampled. The resulting free energy, computed from an adjoint equation, agreed well with the REUS results, illustrating the advantage of accommodating an arbitrary sampling distribution. The committor was computed using the pairwise-distance basis functions described above and indicator functions on the pairwise distances and their time-lagged independent components. The pairwise-distance basis functions gave results that were the least sensitive to lag time, without the artifacts at the boundaries visible in the distance-indicator plots. Reactive fluxes projected onto selected variables allowed visualization of pathways and transition states ($q \approx 0.5$), which were consistent with earlier transition path sampling and shooting studies.
% [ref] fergusonsrv, refs

\begin{figure*}
\includegraphics[width=0.95\textwidth]{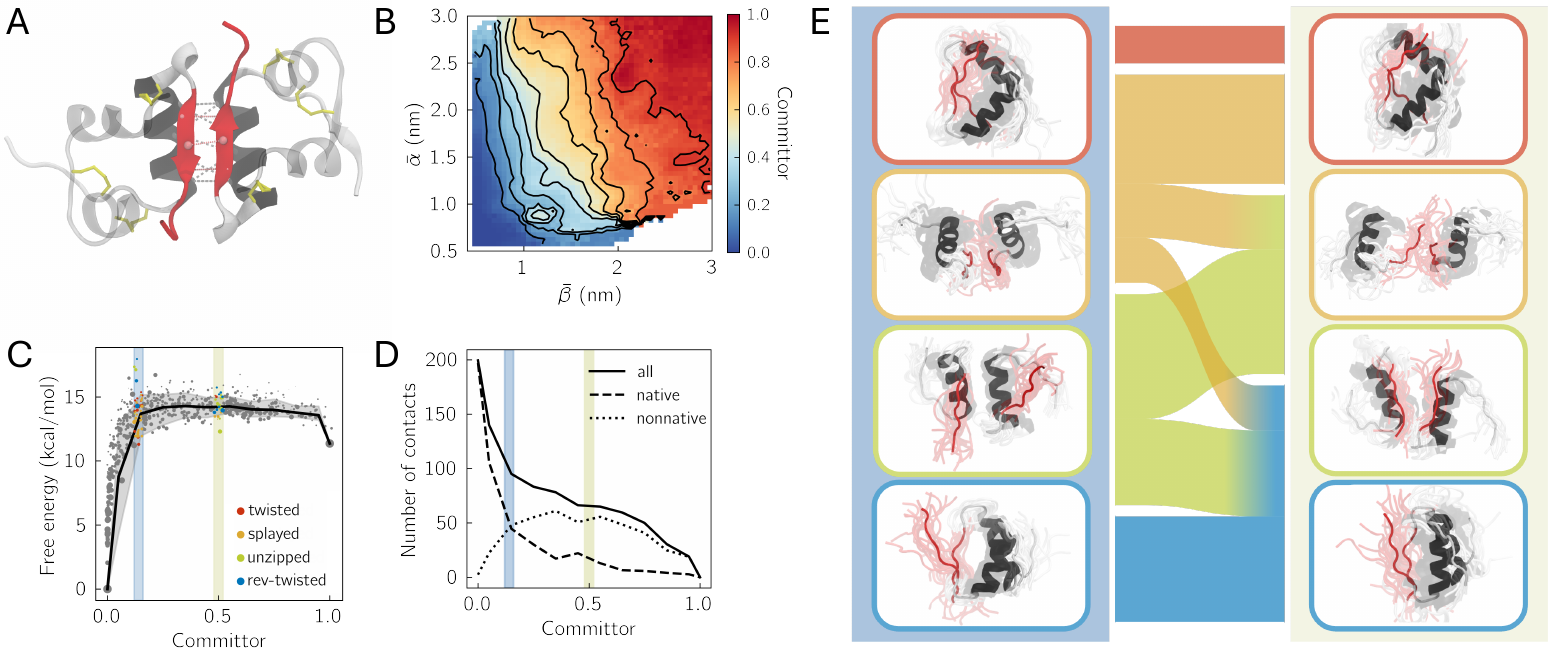}
\caption{
Insulin dimer dissociates through multiple pathways.  (A) Insulin dimer structure.  (B)  The committor  plotted as a function of average distances of selected interfacial $\alpha$-helical contacts ($\bar{\alpha}$) and selected interfacial $\beta$-strand contacts ($\bar{\beta}$); selected contacts are indicated in pink in panel A.  (C) The potential of mean force (PMF) as a function of the committor motivates analyzing the structures at $q\approx 0.14$ (light blue) and $q\approx 0.50$ (light green).  (D) The rapid rise in the PMF reflects loss of native contacts, while $q\approx 0.50$ corresponds to the peak in nonnative contacts.  (E) Representative structures of four pathways (red, orange, green, blue) at $q\approx 0.14$ (light blue) and $q\approx 0.50$ (light green) and the fluxes between them.  Adapted from Ref.\ \citenum{jeong2024analysis}.  Copyright 2024 American Chemical Society. }
\label{fig:insulin}
\end{figure*}

\paragraph{Insulin dimer dissociation.}
The protein hormone insulin forms a homodimer that must dissociate to bind its receptor as a monomer. Beyond its importance for engineering analogs with tailored kinetics, this process serves as a paradigm for coupled (un)folding and (un)binding because the monomer is substantially disordered \cite{antoszewski2020insulin,bustomoner2021structural}; the resulting dynamics are diffusive and multipathway, violating the assumptions of methods that require a well-defined transition state. Jeong {\it et al.}\ \cite{jeong2024analysis} seeded short (5 ns) unbiased trajectories from initial conditions drawn from previous umbrella sampling simulations \cite{antoszewski2020insulin}, for an aggregate of just over 94 $\mu$s, and used an indicator basis set on 45 features (i.e., a stopped MSM) to compute free energies, committors, and reactive fluxes (Fig.\ \ref{fig:insulin}). They identified four dissociation pathways corresponding to rotations around three axes of the dimer structure and quantified, as functions of the committor, both the flux within each pathway and the exchange between them, which they rationalized in terms of the structure. Integrating the flux gave an inverse rate of approximately 100 ms, markedly slower than estimates from Kramers' theory and its extensions \cite{acharya2021rate}; the difference likely reflects the greater resolution of the DGA model, which does not project the barrier onto two dimensions. Characterizing a 100 ms transition with 94 $\mu$s of data illustrates the power of DGA for studying rare events, i.e., ones that occur infrequently compared with thermal fluctuations.

\begin{figure*}
\includegraphics[width=0.95\textwidth]{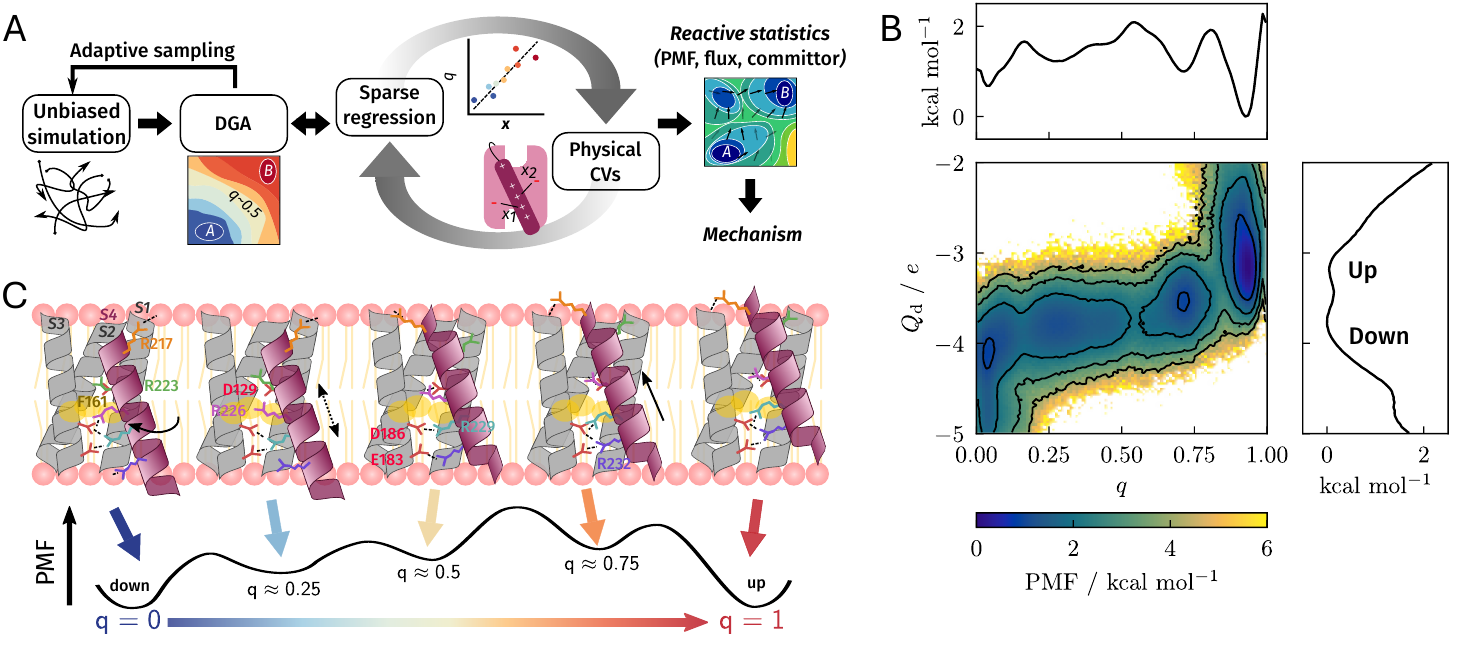}
\caption{
The committor reveals intermediates in the up-down transition of a voltage-sensing domain.  (A) Workflow for the simulations.  (B) Contour plot shows two-dimensional potential of mean force (PMF) as a function of the committor and displacement charge ($Q_d$), which is measurable in electrophysiology experiments.  The one-dimensional PMF for $Q_d$ suggests a two-state mechanism, while that for $q$ suggests there are intermediates that are hidden to $Q_d$, reconciling fluorescence and electrophysiology experiments \cite{villalbagalea2008s4}.  (C) Summary of the mechanism mapped to the committor PMF.    
Adapted from Ref.\ \citenum{guo2024dynamics}, distributed under a Creative Commons Attribution 4.0 International License (\href{http://creativecommons.org/licenses/by/4.0/}{CC BY 4.0}).
}
\label{fig:vsd}
\end{figure*}

\paragraph{Voltage sensing.}
Voltage-sensing domains (VSDs), which underlie processes ranging from pain sensing to cardiac function, respond to changes in transmembrane potential by moving helices bearing positively charged residues. Crystal structures of a phosphatase VSD homologous to voltage-gated ion channel VSDs suggested a helical-screw mechanism in which the S4 helix translocates and rotates to exchange salt-bridge partners  \cite{li2014structural}, but the details of the transition remained unknown. Guo {\it et al.}\ \cite{guo2024dynamics} analyzed 415 $\mu$s of unbiased trajectories initiated from equilibrated REUS windows \cite{shen2022mechanism} and then extended adaptively toward conformations with $q\approx 0.5$ (Fig.\ \ref{fig:vsd}). Because DGA enabled computing committor values for millions of sampled structures, the committor could be used as a reaction coordinate. The free energy as a function of the displacement charge was two-state, consistent with electrophysiological measurements, but as a function of the committor it revealed minima at $q \approx 0.25$, 0.50, and 0.75 (in addition to $q\approx 0$ and $q\approx 1$), indicating five-state microscopic dynamics. Sparse regression of the committor identified nine interpretable physical variables, which revealed a noncanonical helical-screw mechanism in which translation, rotation, and charged side-chain movements of the sensing helix are only loosely coupled. This mechanism explains why intermediates that give rise to complex kinetics in fluorescence measurements are hidden from electrophysiology measurements \cite{villalbagalea2008s4}.

\begin{figure*}
\includegraphics[width=0.95\textwidth]{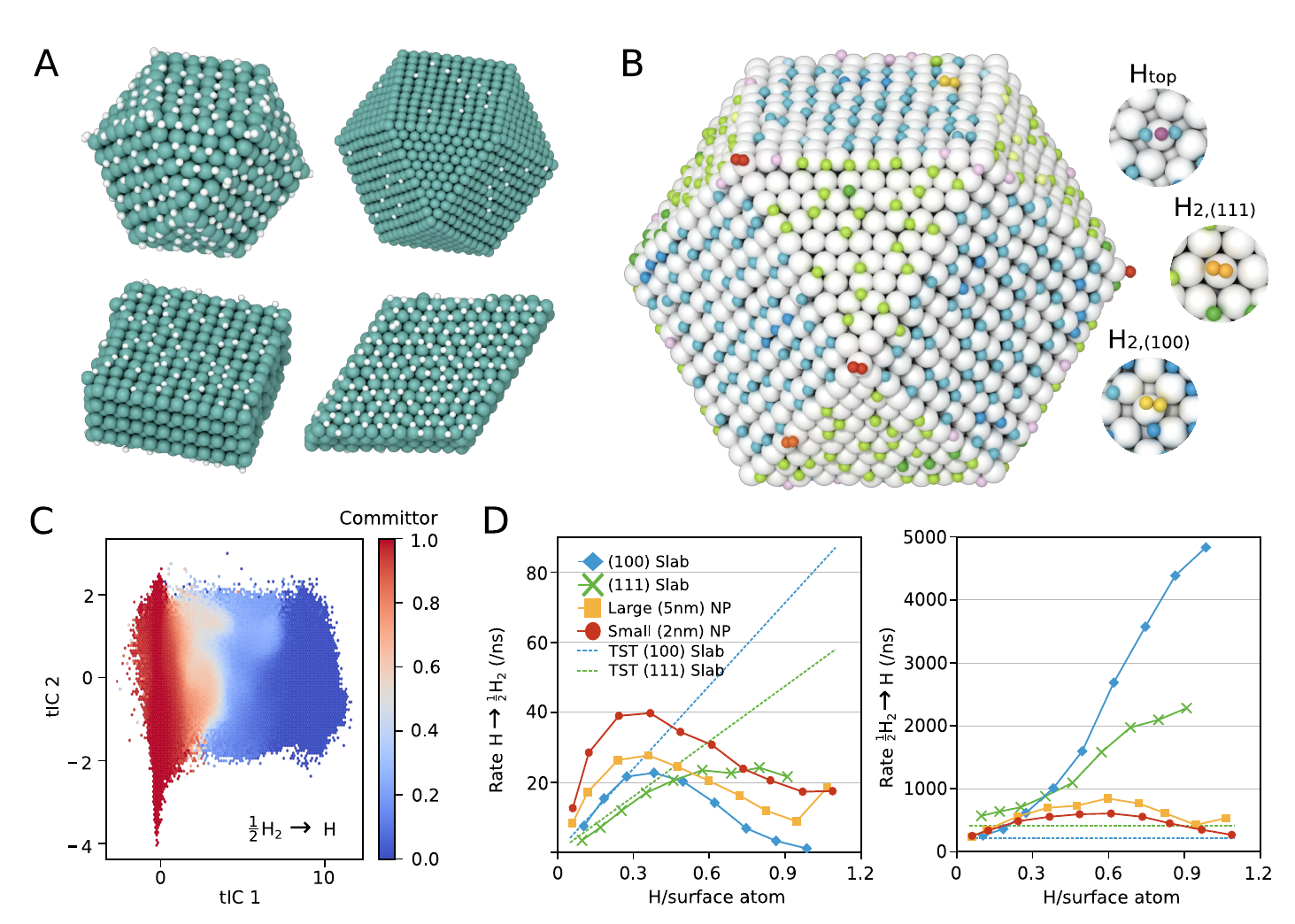}
\caption{
DGA (stopped MSMs) for hydrogen dissociation/association on rhodium. (A) Systems studied:  2 nm nanoparticle, 5 nm nanoparticle, (100) slab, and (111) slab.
%(B) The nearest distance to a neighboring hydrogen atom correlates closely with the 
(B) Simulation snapshot illustrating hydrogen atom and molecule geometric states. Rhodium
atoms are large, white spheres and hydrogen atoms are small spheres: H atoms on (100)/(111)/edges/top sites are blue/green/light purple/dark purple, and  H$_2$ molecules on corners/edges/facets are red/orange/yellow.
(C) Committor for  dissociation. (D) Rates of the association  and dissociation of hydrogen as a function of coverage. Dashed lines
show TST rates on (100) and (111) surfaces.
Adapted from Ref.\ \citenum{mccandler2026markov}.  Copyright 2026 American Chemical Society. 
}
\label{fig:catalysis}
\end{figure*}

\paragraph{Hydrogen dynamics on catalytic surfaces.}
In heterogeneous catalysis, many equivalent reactants and
intermediates interact on a fluctuating surface, so the notion of a
single well-defined transition state that underlies TST is generally
inappropriate.  McCandler {\it et al.}\ \cite{mccandler2026markov} analyzed
association and dissociation of hydrogen on rhodium slabs and cuboctahedral
nanoparticles, using an atomic cluster expansion machine-learned interatomic
potential to generate 200 ns of trajectory data for each of four surface
geometries at hydrogen coverages ranging from about 0.05 to 1.0 monolayer (Fig.\ \ref{fig:catalysis}).
Rather than describing the system as a whole, they represented each hydrogen
atom by 131 features of its local environment within 7 \AA, reduced these to
five time-lagged independent components, and clustered the result into
states.  Because the features are local in space, chemically similar
environments at different locations on the surface are assigned to the same
state, and the dynamics of the full system follow from aggregating over the
individual hydrogen atoms.  The committor for
dissociation was computed by DGA with indicator functions (stopped MSMs), and the rate was obtained from the net reactive flux.  
The resulting rates agreed with TST in the dilute limit but deviated at higher
coverage: trapping at corners and edges of the nanoparticles slowed association
and dissociation, and hydrogen--hydrogen interactions caused the rates to depend
nonmonotonically on coverage.

The results above show that DGA (including stopped MSMs) can yield accurate and informative results for systems of chemical interest.
%, particularly when used together with frameworks for treating memory discussed in Section \ref{sec:nonmarkov}, as the insulin study illustrates. 
However, the expressiveness of these models is ultimately determined by the basis functions and the features that they take as inputs, and choosing them can be difficult for molecules and processes that are less well-characterized than the ones described here.  NNs, which we introduced in the context of solving PDEs (Section~\ref{sec:PDE}), address this issue by learning representations from data.  The remainder of Section \ref{sec:trajectory} focuses on NNs for learning the committor from trajectory data.

\subsection{Variational methods for reversible dynamics}
\label{sec:VCN}

In the \emph{variational committor network} (VCN) approach \cite{chen2023discovering}, one minimizes 
\begin{equation}
  J^{\mathrm{VCN}}_\tau(\theta)
  = \frac{1}{2\tau}
    \E_{\X_0\sim\pi}\!\big[(q_\theta(\X_\tau) - q_\theta(\X_0))^2\big]
  \label{eq:VCN_loss}
\end{equation}
subject to the boundary conditions $q_\theta(\bx) = 0$ for $\bx\in A$ and $q_\theta(\bx) = 1$ for $\bx\in B$. 
%Roux \cite{roux2021string,roux2022transition} derived \eqref{eq:VCN_loss} from the TPT reactive flux rather than from the generator.  The reactive flux is the same through all isocommittor surfaces, so one is free to average over them; this average converts the flux integral from a form that is linear in the committor to one that is quadratic.
This variational principle, which can be derived by considering the TPT reactive flux through a dividing surface \cite{roux2021string,roux2022transition}, has been used with other function approximators as well \cite{banushkina2015nonparametric,aristoff2024fast}. 

Eq.\ \eqref{eq:VCN_loss} reduces to the Dirichlet form \eqref{eq:DF_loss} as $\tau\to 0$.  To see this, note that stationarity of $\pi(\bx)$ implies $\E_\pi[q_\theta^2(\X_\tau)] =\E_\pi[q_\theta^2(\X_0)]$, so that expanding the square in \eqref{eq:VCN_loss} leaves
\begin{equation}
  J^{\mathrm{VCN}}_\tau(\theta)
  = -\frac{1}{\tau}
     \E_{\X_0\sim\pi}\!\big[q_\theta(\X_0)\big(
       q_\theta(\X_\tau) - q_\theta(\X_0)\big)
     \big].
  \label{eq:VCN_DF_identity}
\end{equation}
Conditioning on $\X_0$ and taking $\tau\to 0$, the factor $\tau^{-1}(\E[q_\theta(\X_\tau)\mid \X_0] - q_\theta(\X_0))$ becomes $\calL q_\theta(\X_0)$ by \eqref{eq:generator}, and we recover $J^{\mathrm{DF}}[q_\theta] = -\E_{\X_0\sim\pi}[q_\theta(\X_0)\,\calL q_\theta(\X_0)]$.  In this sense, \eqref{eq:VCN_loss} is the Dirichlet form with the generator replaced by a finite-difference approximation, allowing it to be evaluated from trajectory data without an explicit form for $\calL$.  
% Like \eqref{eq:rate_DF}, \eqref{eq:VCN_DF_identity} should not be used to estimate the committor when the dynamics are not reversible, but once the committor has been estimated \eqref{eq:VCN_DF_identity} can be used to estimate $r_{AB}$ whether or not the dynamics are reversible.
Like \eqref{eq:DF_loss}, \eqref{eq:VCN_DF_identity} should only be minimized to estimate the committor when the dynamics are reversible and the initial states are drawn  from or reweighted to $\pi(\bx)$.

When $\tau$ is finite, trajectory segments can pass through the metastable states $A$ and $B$ or pass all the way from $A$ to $B$ or $B$ to $A$. 
Because it is the finite-difference analog of \eqref{eq:DF_loss}, which applies only as $\tau\to 0$, \eqref{eq:VCN_loss} does not account for these possibilities: it uses the unstopped endpoint $\X_\tau$ rather than the stopped endpoint $\X_{\tau\wedge\T}$, which introduces a systematic error in the committor and rate estimates that grows with $\tau$.

Li {\it et al.}\ \cite{li2022semigroup} introduced a related variational loss that does not suffer from the stopping issue.  They split the average over trajectory segments into two contributions: segments that remain in the transition region for the full lag time, which propagate the trial committor $q_\theta$ forward, and segments that reach $A$ or $B$ within $\tau$, which instead contribute the boundary values at the states they enter. However, the resulting parameter update is not the gradient of any loss function evaluated on finite data (though it is a consistent approximation of the gradient).  Reducing the gradient to a single factor of $\nabla_\theta q_\theta$ requires that the average of $\nabla_\theta q_\theta(\X_0)\,q_\theta(\X_\tau)$ over the data equal the average of $q_\theta(\X_0)\,\nabla_\theta q_\theta(\X_\tau)$, i.e., that it not matter which end of a trajectory segment carries the derivative.  Detailed balance guarantees that they are equal only for an average over infinitely many segments; for a finite data set, the two averages differ. 

The \emph{exact} variational committor network (EVCN) \cite{lorpaiboon2026exact} addresses these issues. Each trajectory segment is divided into pieces at the successive frames at which it is in $A$ or $B$, and the loss sums the squared change in $q_\theta$ over the pieces:
\begin{equation}
  J^{\mathrm{EVCN}}_\tau(\theta)
  = \frac{1}{2\tau}
     \E_{\X_0\sim\pi}\!\Bigg[\sum_{k\geq 0}\big(
       q_\theta(\X_{\tau\wedge\T_{k+1}}) - q_\theta(\X_{\tau\wedge\T_k})\big)^2
     \Bigg],
  \label{eq:EVCN_loss}
\end{equation}
where $\T_0=0$, $\T_{k+1}$ is the first frame after $\T_k$ at which the trajectory is in $A$ or $B$, and $q_\theta(\bx)$ is masked (Section~\ref{sec:PINN}) so that $q_\theta(\bx)=0$ for $\bx\in A$ and $q_\theta(\bx)=1$ for  $\bx\in B$. For $\X_0\notin A\cup B$, $\T_1=\T$, the time of first reaching $A\cup B$ used in \eqref{eq:stopped_operator}; terms with $\T_k\geq\tau$ vanish, so the sum is finite. Segments with $\T>\tau$ contribute only the VCN term in \eqref{eq:VCN_loss}. In the remaining segments, the first and last pieces compare $q_\theta$ at the start or end of the segment with the boundary value of the state reached first or visited last; these terms implement stopping. Each interior piece begins and ends in $A\cup B$, so it contributes 1 if it connects $A$ and $B$ and 0 otherwise. Together, the interior pieces count the complete transitions within the lag time; this count does not depend on $\theta$, but it makes the loss estimate the reactive flux itself, so that its value at the exact committor is $r_{AB}$ for any $\tau$.

EVCN also symmetrizes the empirical loss by evaluating each trajectory segment together with its time reverse, so that the averages of $\nabla_\theta q_\theta(\X_0)\,q_\theta(\X_\tau)$ and $q_\theta(\X_0)\,\nabla_\theta q_\theta(\X_\tau)$ are equal by construction rather than only in the limit of infinite data; the parameter update is then the exact gradient of the loss for any finite data set. Because of stopping and the transition count in \eqref{eq:EVCN_loss}, the EVCN loss can be used both to learn the committor for any $\tau$ when the dynamics are reversible and to compute the rate from a committor estimate regardless of whether the dynamics are reversible. The value of $\tau$ used to estimate the rate need not be the same as that used to estimate the committor.

\begin{figure}
\includegraphics[width=0.475\textwidth]{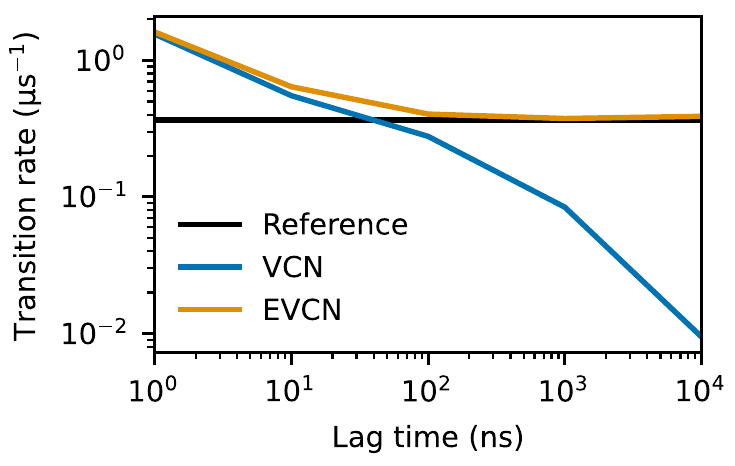}
\caption{
Comparison of folding rate estimates for the 35-residue villin headpiece
subdomain.  While VCN (without stopping) and EVCN (with stopping) give similar estimates at short lag times, only EVCN goes to the empirical reference rate at long lag times.  Reproduced from Ref.\ \citenum{lorpaiboon2026exact}.  Copyright 2026 American Chemical Society. 
}
\label{fig:evcn}
\end{figure}

Lorpaiboon {\it et al.}\ \cite{lorpaiboon2026exact} compared VCN and EVCN loss functions for lag times spanning four orders of magnitude, using data for folding of Trp-cage and villin and for the left-right helix transition of a nine-residue peptide of 2-aminoisobutyric acid (AIB\textsubscript{9}).  The two loss functions perform comparably at the shortest lag times, as expected because the VCN loss is exact for a single time step.  Chen {\it et al.}\ \cite{chen2023discovering} suggest choosing the smallest lag time that renders the dynamics effectively Markovian in the input variables, but care is needed. As $\tau$ increases, the VCN committor flattens in the transition region until its predictions collapse to nearly a single value, and the corresponding rates fall off as $1/\tau$ (Fig.\ \ref{fig:evcn}).  EVCN committors remain accurate at all lag times tested, and their rates plateau near the empirical values computed from the trajectories.  These results illustrate the practical importance of stopping trajectories when they reach $A$ or $B$.

\subsection{Minimizing the fixed-point residual}
\label{sec:FP}

With stopping, the variational methods described in Section \ref{sec:VCN} can yield highly accurate results, as compared with empirical estimates.  However, as already mentioned, they require the initial conditions to be drawn from or reweighted to the stationary distribution $\pi(\bx)$.  Generally $\pi(\bx)$ is large near the metastable states and small near the transition states, which are of greatest interest for learning mechanisms.  Statistically, methods that allow for an arbitrary sample distribution $\mu(\bx)$, as the PINN approach and MSMs/DGA do, enable concentrating sampling near the transition states and, more generally, adaptively sampling (as in DGA examples \cite{strahan2021long,guo2024dynamics} above). Here we discuss approaches that can learn the committor and other statistics of the form in \eqref{eq:general_bvp} from trajectory data without an explicit form for $\calL$ or $\calS$ and without assuming reversibility or $\pi(\bx)$-distributed initial conditions.

\subsubsection{The two-trajectory trick}
\label{sec:two_traj}

The direct finite-$\tau$ analog of the PINN approach is to minimize the squared residual of \eqref{eq:fixed_point},
\begin{equation}
  J^{\mathrm{FP}}
  = \E_{\X_0\sim\mu}\!\big[
      (\mathcal{S}^\tau q_\theta(\X_0) - q_\theta(\X_0))^2
    \big],
  \label{eq:FP_loss}
\end{equation}
with boundary conditions treated as described in Section \ref{sec:PINN}. One way to minimize \eqref{eq:FP_loss} is to follow its gradient:
\begin{equation}\label{eq:gradJFP}
\begin{split}
  \nabla_\theta J^{\mathrm{FP}}
  = 2\E_{\X_0\sim\mu}\big[
    &\mathcal{S}^\tau q_\theta\,\mathcal{S}^\tau \nabla_\theta q_\theta
     - \mathcal{S}^\tau q_\theta\,\nabla_\theta q_\theta\\
    &- q_\theta\,\mathcal{S}^\tau \nabla_\theta q_\theta
     + q_\theta\,\nabla_\theta q_\theta\big],
\end{split}
\end{equation}
where we suppress the argument of $q_\theta(\X_0)$ for clarity and use $\nabla_\theta \mathcal{S}^\tau q_\theta = \mathcal{S}^\tau \nabla_\theta q_\theta$ because $\calS$ is linear.  
%The last three terms on the right-hand side of \eqref{eq:gradJFP}, which each have zero or one instance of $\calS$, which is itself an expectation (cf.\ \eqref{eq:stopped_operator}), can be computed from trajectories by means analogous to \eqref{eq:Ctau}.  
Each factor of $\calS$ in \eqref{eq:gradJFP} corresponds to an expectation conditioned on $\X_0$ (see \eqref{eq:stopped_operator}).
Because each of the last three terms on the right-hand side of \eqref{eq:gradJFP} contains at most one such expectation, they can be estimated from trajectory data by means analogous to \eqref{eq:Ctau}.
By contrast, because the first term involves a product of two expectations conditioned on the same $\X_0$, it cannot be estimated consistently from one trajectory per initial condition; in general, for functions $f$ and $g$ 
\begin{multline}
    \E[f(\X_{\tau\wedge \T})\mid \X_0]\,\E[g(\X_{\tau\wedge \T})\mid \X_0]\neq\\ \E[f(\X_{\tau\wedge \T})g(\X_{\tau\wedge \T})\mid \X_0].
\end{multline}
% Instead, \cite{mitchell2024committor} uses a ``swarm'' of many trajectories per intial condition to estimate $\mathcal{S}^\tau q_\theta(X_0^{(m)})$ and $\mathcal{S}^\tau \nabla q_\theta(X_0^{(m)})$ for every $X_0^{(m)}$. Any remaining variance in those estimates will lead to a bias in the corresponding estimate of \eqref{eq:gradJFP}.

The \emph{two-trajectory trick}  resolves this issue by generating two independent trajectories from each $\X_0^{(m)}$, with endpoints $\X_{\tau\wedge \T^{(m,1)}}^{(m,1)}$ and $\X_{\tau\wedge \T^{(m,2)}}^{(m,2)}$, and computing %the first term on the right-hand side of \eqref{eq:gradJFP}
$\E_{\X_0\sim\mu}[\mathcal{S}^\tau q_\theta\,\mathcal{S}^\tau \nabla_\theta q_\theta]$
as \cite{strahan2023predicting,qu2024deep,qu2026deep}
\begin{equation}
    \E_{\X_0\sim\mu}[q_\theta(\X_{\tau\wedge \T^{(m,1)}}^{(m,1)})\,
      \nabla_\theta q_\theta(\X_{\tau\wedge \T^{(m,2)}}^{(m,2)})].
\end{equation}
Generating more than two endpoints per initial condition and averaging over distinct pairs can reduce the variance \cite{strahan2023predicting}.
The two-trajectory trick thus enables computing the gradient in \eqref{eq:gradJFP} at the expense of requiring multiple trajectories for each initial condition. Most available molecular dynamics data sets contain only one trajectory from each initial condition.

\subsubsection{Fixed-point iteration}
\label{sec:richardson}

An alternative way to solve \eqref{eq:fixed_point} that avoids the problematic term in \eqref{eq:gradJFP} is the iteration \cite{strahan2023inexact}
\begin{align}
  q_{\theta_{k+1}}(\bx)
  &= q_{\theta_{k}}(\bx)
     + \eta\,(\mathcal{S}^\tau q_{\theta_{k}}(\bx) - q_{\theta_{k}}(\bx)),
  \label{eq:richardson}
\end{align}
where $q_{\theta_k}(\bx)$ is the model for the committor with parameters $\theta_k$ at iteration $k$.  
The hyperparameter $\eta$ controls the rate at which the model changes; $\eta\in (0,1]$ guarantees convergence of the exact iteration \eqref{eq:richardson}. However,  
because $q_{\theta_k}(\bx)$ is a parameterized function (e.g., an NN), one cannot directly assign it values as \eqref{eq:richardson} suggests.  Instead, the parameters must be adjusted to make the model as close as possible to the right-hand side of \eqref{eq:richardson}.  Denoting the latter by $\tilde{q}$, 
\begin{equation}
  \theta_{k+1} = \operatorname{arg\,min}_\theta\, d(\tilde{q},\, q_\theta),
  \label{eq:projection}
\end{equation}
where $d$ is a measure of dissimilarity, the choice of which is discussed below. 

In practice, the target $\tilde{q}$ is not available as an explicit function;  instead, for each initial condition $\X_0^{(m)}$, one obtains a random sample $\tilde{q}^{(m)}$ of $\tilde{q}(\X_0^{(m)})$:
%$\E[\tilde{q}^{(m)}\mid \X_0^{(m)}] = \tilde{q}(\X_0^{(m)})$. 
%
\begin{equation}
  \tilde{q}^{(m)}
  = (1-\eta)\,q_{\theta_{k}}(\X_0^{(m)})
    + \eta\, q_{\theta_{k}}(\X_{\tau\wedge \T^{(m)}}^{(m)}).
  \label{eq:noisy_target}
\end{equation}
%
%Masking (Section~\ref{sec:PINN}) is the preferred way to treat the boundary conditions. When the stopped endpoint in \eqref{eq:noisy_target} lies in $A$ or $B$, any error in the boundary values enters the targets and carries over to the converged solution.
Masking, which again can be combined with boundary penalty terms to mitigate discontinuities at the boundaries (Section~\ref{sec:PINN}), prevents errors in $A$ and $B$ from propagating to the rest of the solution.

The samples $\{\tilde{q}^{(m)}\}$ are unbiased but noisy.
When one takes advantage of the fact that \eqref{eq:noisy_target} requires only one trajectory per initial condition, one cannot average the noise pointwise and must instead treat it through the choice of $d$ used for the regression in \eqref{eq:projection}. If $\nabla_\theta d(\tilde{q}, q_\theta)$ is linear in $\tilde{q}$, the errors in the individual targets average to zero over the data distribution, and the samples $\{\tilde{q}^{(m)}\}$ give an unbiased estimate of the gradient.
Two standard losses that satisfy this condition are least-squares and binary cross-entropy,
\begin{align}\label{eq:cross_entropy}
    d(\tilde{q},q_\theta)
  = -\E_{\X_0\sim\mu}\!\big[&
      \tilde{q}(\X_0)\ln q_\theta(\X_0) \\
      &+ (1-\tilde{q}(\X_0))\ln(1-q_\theta(\X_0))
    \big], \nonumber
\end{align}
but the latter is the natural choice for a probability such as the committor.
When the output of the network is constrained to $[0,1]$ by a sigmoid, the gradient of a least-squares loss vanishes as $q_\theta$ approaches $0$ and $1$, whereas that of the binary cross-entropy does not. Ref.~\citenum{strahan2023inexact} uses this construction and terms \eqref{eq:cross_entropy} the softplus loss. 
More generally, the $d$ functions with the required linearity properties for $\nabla_\theta d(\tilde{q}, q_\theta)$ are called Bregman divergences \cite{Gneiting2011}. This requirement  can be relaxed if one generates a swarm of trajectories per initial condition $\X_0^{(m)}$ and replaces $\tilde q^{(m)}$ by a sample average approximation of $\tilde q(\X_0^{(m)})$ as in Ref.\ \citenum{mitchell2024committor}. However, any remaining variance in those estimates leads to a bias in the corresponding estimate of $\nabla_\theta d(\tilde{q}, q_\theta)$. 
When $\eta$ is small, \eqref{eq:richardson} approximates an ordinary differential equation; changing variables in that equation offers further flexibility. 

\begin{figure*}
\includegraphics[width=0.95\textwidth]{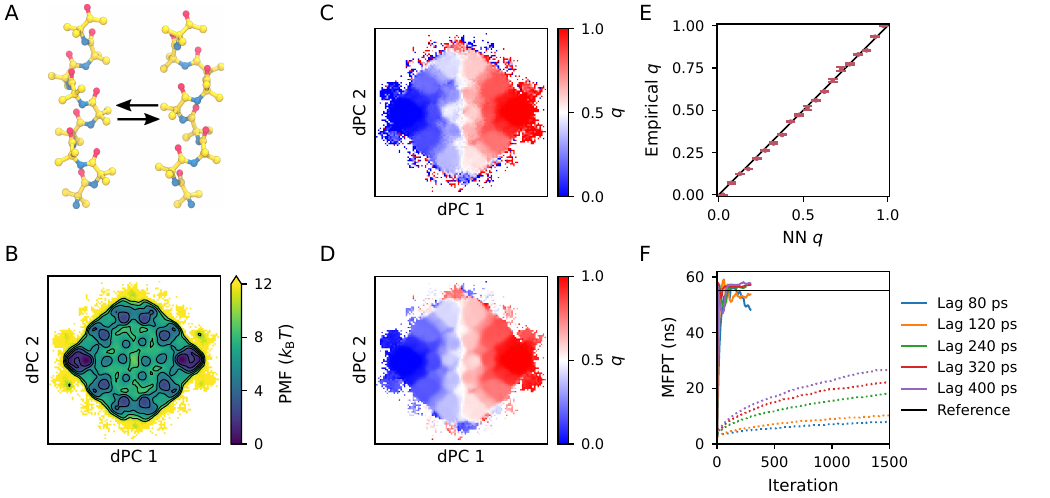}
\caption{
Fixed-point iteration yields accurate committors for the AIB$_9$ left-right helix transition.  (A) Left- and right-helix structures. Carbon, nitrogen, and oxygen atoms are shown in yellow, blue, and red, respectively; hydrogen atoms are omitted for clarity. (B) Potential of mean force constructed from the histogram of value pairs of the first two dihedral angle principal components (dPCs); data are from 20 trajectories of 5 $\mu$s. (C) Averages of $\chi_B(\X_\T)$ (empirical committor) for initial conditions in bins in the first two dPCs computed from the same 20 trajectories. (D) Averages of representative neural-network committors trained on the data set of 6,910 trajectories of 20 ns; $\tau$ corresponds to 400 ps. (E) Comparison between the empirical committor and the neural-network committor (trained as in panel D). Error bars indicate standard deviations over ten different initializations of the neural-network parameters.  (F) MFPT estimate convergence without (dotted lines) and with (solid lines) subspace acceleration ($p=5$). Adapted with permission from Ref.\ \citenum{strahan2023inexact}.  Copyright 2023 AIP Publishing.}
\label{fig:aib9committor}
\end{figure*}

Fig.\ \ref{fig:aib9committor} shows  committor estimates for the AIB$_9$ left-right helix transition.  
Excellent agreement is obtained between estimates from fixed-point iteration with many short trajectories and empirical references from long trajectories.

\subsection{Subspace acceleration}
\label{sec:subspace}

Just as the eigenvalues of the transition operator $\mathcal{T}^\tau$ give relaxation time scales \cite{swope2004describinga,prinz2011markov,bowman2013introduction}, the eigenvalues of the stopped transition operator $\mathcal{S}^\tau$ control the rate of convergence of the neural-network methods discussed in Sections~\ref{sec:VCN} and~\ref{sec:FP} \cite{strahan2023inexact}. Because the boundary conditions are enforced separately (e.g., by masking, Section~\ref{sec:PINN}), the part of $q_\theta$ that is adjusted during training vanishes on $A\cup B$, and the relevant space consists of functions with this property. On this space, $\mathcal{S}^\tau$ has no eigenvalue exactly equal to 1 because each application of $\mathcal{S}^\tau$ transfers probability to $A\cup B$. When the time to reach $A\cup B$ is long, as often occurs for the molecular transitions considered here, the dominant eigenvalue $\lambda_0$ approaches 1, and convergence becomes slow.

In this case, convergence can be accelerated by an iteration that also learns the dominant $p$-dimensional eigenspace \cite{strahan2023inexact}, in the spirit of block power iteration. This procedure shifts control of the rate of convergence to the gap between $\lambda_{p-1}$ and $\lambda_p$. Operationally, one initializes $p$ NNs $\{\phi^\ell_\theta\}_{\ell=1}^p$ and trains them through a nested iteration, repeated until convergence. The inner loop corresponds to the procedure described in Section~\ref{sec:richardson}, in which noisy targets in \eqref{eq:noisy_target} are generated and NNs are trained on them. 
The outer loop executes the inner loop and then ensures that the resulting functions do not become rank deficient, e.g., by QR decomposition.
%to prevent their empirical Gram matrix from becoming rank deficient.
%The outer loop executes the inner loop and then orthogonalizes the resulting functions, e.g., by QR decomposition.
%One obtains the dominant eigenvalues and eigenfunctions of $\mathcal{S}^\tau$ by solving the $p\times p$ generalized eigenvalue problem $C^\tau W = C^0 W\Lambda$, where the matrices $C^t_{k\ell} = \E_{\X_0\sim\mu}[\phi^k_\theta(\X_0)\,\phi^\ell_\theta(\X_{t\wedge \T})]$ (for $t=0,\tau$) are estimated from the trajectory data. 

The trained networks $\{\phi^\ell_\theta\}_{\ell=1}^p$, together with the result of the fixed-point iteration described in Section~\ref{sec:richardson}, can be used as a basis for DGA to compute kinetic statistics.
%Because the eigenfunctions obtained from this procedure span the slow subspace of $\mathcal{S}^\tau$, they are also expected to be effective basis functions for kinetic statistics. 
For the AIB$_9$ left-right helix transition, subspace acceleration with $p=5$ converges to the correct MFPT orders of magnitude faster than the basic fixed-point iteration in Section \ref{sec:richardson} across all values of $\tau$ tested \cite{strahan2023inexact} (Fig.\ \ref{fig:aib9committor}F).

\section{Beyond the Markov Assumption}
\label{sec:nonmarkov}

The methods described in Sections~\ref{sec:PDE}--\ref{sec:trajectory} assume that $\X_t$ is Markovian, i.e., that the future state depends only on the current state. This is the case by construction at each step of an atomic-resolution molecular dynamics simulation, when all positions and velocities are available. However, we generally compute statistics in terms of a reduced set of  variables $\boldsymbol{\xi}(\X_t)\in\mathbb{R}^m$, and the coupling of these variables to the orthogonal ones that are treated implicitly makes the projected process depend on the history, i.e., there is memory \cite{zwanzig2001nonequilibrium,mori1965transport,dalton2025memory}. When the dynamics have a separation of time scales, one can in principle choose $\boldsymbol{\xi}$ that relax slowly compared with the orthogonal variables, so that the memory decays quickly and can be neglected. However, not all systems have a clear separation of time scales, and, even when they do, finding variables that minimize the memory can be challenging \cite{guttenberg2013minimizing}. 
% [ref] zwanzig_book, mori1965
Discretizing the state space in an MSM and expanding a statistic in a finite basis in DGA \eqref{eq:DGA_ansatz} are also projections that introduce memory, even when $\boldsymbol{\xi}$ are well chosen.

One way to test whether the memory is significant is to vary the lag time $\tau$. In MSMs, the implied relaxation time scales $t_k(\tau) = -\tau/\ln\lambda_k(\tau)$, where  $\lambda_k$ are the eigenvalues of the transition matrix, should plateau once the memory decays \cite{swope2004describinga,prinz2011markov,bowman2013introduction}.  Because \eqref{eq:fixed_point} is valid for any $\tau>0$ when the dynamics are Markovian, the committor should be similarly independent of $\tau$; other statistics based on the stopped transition operator $\calS$ should be as well. Extending $\tau$ until the memory decays comes at the statistical cost of decreasing the number of independent trajectory segments obtainable from a given amount of data.  For this reason, it can be useful to account explicitly for history.

\subsection{Mori--Zwanzig formalism}

The standard approach for treating memory in chemical physics is the Mori--Zwanzig formalism \cite{mori1965transport,zwanzig2001nonequilibrium,dalton2025memory}. Projecting functions of the full state onto functions of the reduced set of variables $\boldsymbol{\xi}$ separates the dynamics exactly into three contributions: a Markovian term that depends only on the current values of $\boldsymbol{\xi}$, a memory term that integrates the history of $\boldsymbol{\xi}$ against a kernel, and a term that reflects the dynamics of the orthogonal variables. Building on quasi-MSMs \cite{cao2020advantages,dominic2023building,dominic2023memory}, Lorpaiboon {\it et al.}\ \cite{lorpaiboon2024accurate} used Mori--Zwanzig decompositions to reduce the error that the basis projection introduces to DGA. They derived explicit expressions for the forward and backward committors, the MFPT, and the stationary distribution in terms of analogs of the matrices and vectors in \eqref{eq:Ctau} and \eqref{eq:bvec}, which can be estimated from trajectory data. Because these expressions determine the memory at each lag from the correlation matrices together with the memory at shorter lags, the kernel is obtained recursively from data without assuming a functional form. Here the decomposition takes the form of a discrete-time generalized master equation, in which the memory term is a sum over times spaced by a short lag $\shortlag$ within a window $\tau$, with $\tau/\shortlag$ a positive integer; taking $\tau/\shortlag=1$ recovers standard DGA. The orthogonal-variable term is neglected, but each term in the memory sum applies the complementary projection again, which removes the dynamics that the basis represents; when the basis captures most of the slow dynamics, the neglected remainder is smaller than the one standard DGA neglects. 

It is worth contrasting quasi-MSMs and DGA with memory with traditional Mori--Zwanzig analyses \cite{zwanzig2001nonequilibrium,dalton2025memory}. In the latter, the projection is typically onto a low-dimensional space, so that it is hard to capture all the slow dynamics and, in turn, to determine a memory kernel that yields accurate statistics. Many studies assume a functional form for the kernel, typically exponential. Studies that use more flexible parameterizations indicate that no simple universal form exists \cite{perico1993positional,chorin2002optimal,lei2016data,grogan2020data,ayaz2021non,vroylandt2022likelihood}. The approach above determines the kernel from the data without such an assumption.
Because quasi-MSMs and DGA with memory project onto many states or basis functions, which are often functions of many variables, they can retain most of the slow dynamics. The memory then decays rapidly, and few terms suffice, with most of the correction coming from the first term beyond the Markov approximation.

For a two-dimensional triple-well potential and the left-right helix transition of AIB$_9$, memory reduced the lag time needed to match reference MFPTs and stationary distributions by about an order of magnitude \cite{lorpaiboon2024accurate}. Jeong {\it et al.}\ \cite{jeong2024analysis} employed DGA with memory in the insulin dimer dissociation study discussed above and found $\tau/\shortlag=2$, a single term beyond the Markov approximation, to be sufficient. Strahan  {\it et al.}\  \cite{strahan2023inexact} extended the Mori--Zwanzig approach \cite{darve2009computing,cao2020advantages} to the subspace acceleration described above; there, they applied it after the basis was learned, but it may also be possible to speed convergence by incorporating memory at each (outer loop) iteration.

%DGA with memory has numerical advantages over quasi-MSMs and other methods that predict long-time statistics by integrating the GME. Integrating the GME requires estimating the full coarse-grained propagator and then extrapolating it forward in time. Solving directly for individual kinetic statistics allows better control of errors. In particular, because the boundary conditions are satisfied regardless of the choice of $\tau$ and $\shortlag$, one can increase $\tau$ to decrease the error while keeping $\tau/\shortlag$ small.

\subsection{Delay embedding}

Delay embedding is an alternative way to account for memory that is widely used in nonlinear dynamics. One augments the state with a window of $n$ previous observations, $\tilde{\X}_t = (\X_t, \X_{t-\shortlag}, \X_{t-2\shortlag},\ldots, \X_{t-n\shortlag})$ where $\shortlag$ is again a short lag, and then applies any of the estimators of Section~\ref{sec:trajectory} unchanged. The motivation is Takens' theorem \cite{takens1981detecting,stark1997takens}, which states that a sufficiently long window of observations generically reconstructs the underlying state; once the window spans the memory time, the augmented process is approximately Markovian. The cost is that the state grows to $m(n+1)$ components for $\boldsymbol{\xi}(\X_t)\in\mathbb{R}^m$, offsetting the advantages of working with a reduced set of variables. Thiede {\it et al.}\ \cite{thiede2019galerkin} and Strahan {\it et al.}\ \cite{strahan2021long} used delay embedding to reduce the number of variables needed to describe Fip35 WW domain and Trp-cage folding, respectively, to a few physically motivated variables. Delay embedding and the Mori--Zwanzig approach thus address the memory differently: the former enlarges the state until the dynamics are approximately Markovian, while the latter leaves the state unchanged and corrects the estimator. It would be interesting to compare the two with regard to data requirements and robustness to hyperparameter choices for a suite of high-dimensional models.

\subsection{Augmented TPT}

History can also be used to extend TPT to statistics that depend on a sequence of events. Traditional TPT applies to trajectories connecting $A$ and $B$, and it computes statistics by factoring the selection of reactive trajectories into quantities that are local in time. This factorization fails for statistics that depend on a sequence of events, such as the reactive current for only those trajectories that pass through an intermediate. Lorpaiboon {\it et al.}\ \cite{lorpaiboon2022augmented} showed that such statistics become accessible by augmenting the state with a label ${\bf Y}_t$ that records the sequence of events in a trajectory, generalizing earlier history-augmented approaches, in which states are labeled by the last metastable state visited \cite{dickson2009separating,vandeneijnden2009exact,suarez2016accurate,vani2022computing}. Provided the augmented process ${\bf Z}_t = (\X_t, {\bf Y}_t)$ is Markovian, TPT applies to it unchanged. Lorpaiboon {\it et al.}\ \cite{lorpaiboon2022augmented} used this framework to resolve the reactive density, reactive current, committor, and conditional MFPTs for individual steps of a reaction through an intermediate and for individual pathways in a multipathway reaction. The framework can also treat reactions that traditional TPT cannot, such as cycles in oscillators and excitable systems, in which the reactant and product states coincide.
Whereas the treatments above use history to address the memory that projection introduces, here the process is already Markovian and history enlarges the class of problems that can be addressed. Because augmented TPT also casts statistics in terms of quantities local in time, the estimators of Section~\ref{sec:trajectory} extend to it.

\begin{table*}
\centering
\caption{Self-supervised methods for computing kinetic statistics.  Methods that do not require an explicit form for the generator $\calL$ compute statistics directly from trajectory data.  Methods that require reversibility also assume that $\X_0$ is drawn from the stationary distribution $\pi(\bx)$; others can accommodate an arbitrary sample distribution $\mu(\bx)$. }
\label{tab:methods}
\small
\setlength{\tabcolsep}{5pt}
\renewcommand{\arraystretch}{1.3}
\begin{tabular}{l l c c c l}
\hline\noalign{\vskip 3pt}
\textbf{Method} & \textbf{Section} & \textbf{Key Eq.}
  & \parbox[b]{2cm}{\centering\textbf{Require\\explicit $\calL$}}
  & \parbox[b]{2cm}{\centering\textbf{Require\\reversibility}}
  & \textbf{Data per $\X_0$} \\ \hline
PINNs                    & \ref{sec:PINN}      & \eqref{eq:PINN_loss}       & \checkmark &            & $\X_0$ \\
Dirichlet form          & \ref{sec:DF}        & \eqref{eq:DF_loss}         & \checkmark & \checkmark & $\X_0$ \\
DGA / stopped MSMs                    & \ref{sec:DGA}       & \eqref{eq:DGA_linear}      &            &            & $(\X_0,\X_{\tau\wedge \T})$ \\
(E)VCN                     & \ref{sec:VCN}      & \eqref{eq:VCN_loss}, \eqref{eq:EVCN_loss}        &            & \checkmark & $(\X_0,\X_{\tau\wedge \T})$ + crossings \\
Two-trajectory trick         & \ref{sec:two_traj}  & \eqref{eq:FP_loss}         &            &            & $(\X_0,\X_{\tau\wedge \T^{(1)}}^{(1)},\X_{\tau\wedge \T^{(2)}}^{(2)})$ \\
Fixed-point iteration   & \ref{sec:richardson}& \eqref{eq:richardson}      &            &            & $(\X_0,\X_{\tau\wedge \T})$ \\ \hline
\end{tabular}
\end{table*}

\section{Choosing the sample distribution}
\label{sec:sampling}

The methods discussed in Sections \ref{sec:PDE} and \ref{sec:trajectory} and their requirements are summarized in Table \ref{tab:methods}.  Except for the variational methods in Sections \ref{sec:DF} and \ref{sec:VCN}, one is free to choose the distribution $\mu$ of initial conditions.  In this section, we consider this choice, which is arguably as important as the estimator \cite{pigeon2026approximating}.

\subsection{Mathematical analysis}
\label{sec:guarantees}

Existing error bounds for MSMs 
\cite{singhal2005error,sarich2010approximation,prinz2011efficient,djurdjevac2012estimating,webber2021error,tuchkov2025error} 
were generally obtained from classical perturbation theory and are controlled by global quantities, such as  condition numbers or escape times that grow exponentially with the barrier height in metastable systems \cite{thiede2015sharp}.
Moreover, kinetic statistics can range over orders of magnitude, making relative, rather than absolute, error the relevant metric.  Existing bounds, therefore, do not address whether statistics such as the committor can be practically estimated to a given relative accuracy for metastable systems.

When addressing this question, it is useful to distinguish two sources of error. Statistical error results from estimating transition probabilities from finite data. Approximation error remains even with infinite data because the basis is finite. 
Recently, Cheng and Weare analyzed statistical error for stopped MSMs on finite state spaces \cite{cheng2024surprising}. They bound the relative asymptotic variance directly in terms of local connectivity properties of the dynamics, without explicit dependence on condition numbers or expected escape times.  

In the case of the committor, the bound takes the form
\begin{equation}
\label{eq:avar_bound}
\lim_{M\to\infty}\frac{M\,{\rm Var}(\hat{q}_i)}{q_i^2}
  \leq \sum_{j,k;\,k\neq j}
  \frac{S^\tau_{jk}}{\mu_j\,\big(Q^\tau_{jk}\big)^2},    
\end{equation}
where $\hat q_i$ is the stopped-MSM estimate of the committor at state $i$ obtained from $M$ trajectory segments, $i$ is any state outside $A\cup B$, $j$ runs over states outside $A\cup B$, $k$ runs over all states, %$S^\tau_{jk}$ are the elements of the stopped transition matrix, 
and $\mu_j$ is the probability of state $j$ in the distribution of initial conditions. In this finite-state-space setting, $\mathcal{S}^\tau$ becomes a matrix whose $jk$ entry, $S^\tau_{jk}$, is the probability that a trajectory launched from state $j$ is in state $k$ at time $\tau\wedge\T$. The quantity $Q^\tau_{jk}$ in the denominator is the probability that a trajectory launched from $j$, observed at intervals of $\tau$, reaches $k$ before returning to $j$ and before reaching any state in $A\cup B$ other than $k$.  
The ratio $S_{jk}^\tau/(Q^\tau_{jk})^2$ thus compares the probability of a direct transition from $j$ to $k$ within one lag time with the probability that the trajectory finds its way from $j$ to $k$ by any route. Because a direct step is one such route, $Q^\tau_{jk}\geq S^\tau_{jk}$, and greater indirect connectivity lowers the bound.
Notably, \eqref{eq:avar_bound} does not depend on the expected escape time, which can be long when transitions are rare compared with thermal fluctuations.

Three conditions make the sum in \eqref{eq:avar_bound} small. (1) No $\mu_j$ in the denominator is too small. (2) Transitions between neighboring states remain probable as the state space is refined, so that a graph retaining only transitions above a fixed probability threshold stays connected. (3) Transition probabilities decay at least as fast as a Gaussian function of the graph distance between states, so that long jumps within one lag time are strongly suppressed. 

When trajectory data are simulated, condition 1 is under user control, and it has immediate implications for the choice of sampling distribution $\mu(\bx)$.  In particular, the common choice $\mu(\bx)\propto\pi(\bx)$ leads to $\mu_j$ that shrink exponentially with barrier height, which inflates the relative asymptotic variance; \eqref{eq:avar_bound} instead suggests that one should choose $\mu(\bx)$ to avoid assigning very small probabilities to dynamically important states.
Conditions 2 and 3 describe locality of the dynamics, which, for small to intermediate $\tau$, is reasonable to assume for molecular transitions with diffusive barrier crossings, the focus of this review.

Under the conditions above, Cheng and Weare show that the relative asymptotic variance grows no faster than the cube of the number of states $N$. One factor of $N$ arises because spreading initial conditions over more states leaves each with a smaller share of the data, and two arise from counting pairs of states in the sum. In the model that they consider, the barrier height grows in proportion to $N$, because the landscape is scaled so that neighboring states differ by about $k_BT$. The bound therefore grows as the cube of the barrier height in units of $k_BT$ and, because the committor decreases exponentially with that barrier, as the cube of $\ln[1/q(\bx)]$.

The relative asymptotic variance sets the computational cost, because the number of trajectories needed for a given relative accuracy is proportional to it. For shooting at a single configuration, the Bernoulli variance $q(1-q)/M$ from Section~\ref{sec:roadmap} gives a relative asymptotic variance of $[1-q(\bx)]/q(\bx)$, which grows exponentially with the barrier height. The cost of the stopped MSMs discussed here instead scales polynomially. The advantage in aggregate simulation time is larger still, because each shooting trajectory runs until reaching $A$ or $B$.

The advantage also depends on the lag time. As $\tau$ grows, a trajectory segment long enough to reach $A$ or $B$ ends there, so the direct step becomes the only route from $j$ to $k$ that contributes,  $Q^\tau_{jk}$ approaches $S^\tau_{jk}$, and the summand approaches $1/(\mu_j S^\tau_{jk})$, which is large for rare transitions. The relative asymptotic variance behaves the same way, approaching that of shooting as $\tau\to\infty$. In numerical tests, the relative asymptotic variance is smallest at an intermediate lag time. The lag-time dependence compounds the statistical cost of long lag times noted in Section~\ref{sec:nonmarkov}, and it strengthens the case for correcting the estimator for memory rather than extending $\tau$ until the dynamics appear Markovian.

\subsection{Sampling strategies}
\label{sec:samplingpractice}

A central result of the previous section is that the choice $\mu(\bx)\propto \pi(\bx)$ is inefficient and that $\mu(\bx)$ should instead cover the transition region well.  Distributing initial points in the latter way generally requires some knowledge of variables that can describe the transition. This might appear to present a chicken-and-egg problem because the best such variables are those that correlate with the committor. However, sampling can be initiated using variables that are sufficient for separating $A$ and $B$; they need not specify the full mechanism.  The results can then be used to guide adaptive sampling. 

When one has access to previous data for a system, one can draw initial points from them. For example, DGA studies draw initial points from earlier umbrella sampling \cite{guo2024dynamics,jeong2024analysis} and temperature replica exchange \cite{zhang2024temperature} simulations.  If no such data are available, reasonable initial points can be obtained by repeatedly gently steering the system through the transition, e.g., with adiabatic bias molecular dynamics \cite{marchi1999adiabatic}, as in Refs.\ \citenum{strahan2021long,antoszewski2021kinetics}. If comparisons with earlier simulations indicate that the regions of configuration space that contribute to the reaction are not sampled adequately, additional initial points can be added \cite{strahan2021long,guo2024dynamics}.

Methods that sample unbiased trajectories as a means of exploring the state space, such as adaptive search methods \cite{kleiman2023adaptive,wang2018constructing,pengmei2026recursive}, the string method with swarms of trajectories \cite{pan2008finding}, and steered transition path sampling \cite{guttenberg2012steered}, could alternatively be used to obtain not only initial points but also an initial sample of trajectories. Some studies bias and correct the dynamics  \cite{sidky2020machine,shmilovich2023girsanov,donati2022review,keller2024dynamical,chen2023discovering,megias2025iterative}, but because the variance of the reweighting factor grows exponentially with $\tau$, care is needed when pursuing this strategy.

More systematic sampling can be achieved through splitting and trajectory stratification algorithms such as weighted ensemble \cite{huber1996weighted,zuckerman2017weighted,aristoff2023weighted}, forward flux sampling \cite{allen2006simulating,allen2009forward}, transition interface sampling \cite{vanerp2003novel,hall2022practical}, milestoning \cite{faradjian2004computing,bellorivas2015exact,elber2020milestoning}, and nonequilibrium umbrella sampling \cite{warmflash2007umbrella,dickson2010enhanced,dinner2018trajectory}.  These algorithms divide a space of progress variables into regions and launch relatively short unbiased molecular dynamics simulations so as to converge estimates of rates computed directly from sampled fluxes. While it is possible to modify these algorithms to obtain point estimates of the committor \cite{vandeneijnden2009exact,vani2022computing}, the trajectory segments that they sample can be used with the estimators that we describe here to obtain models of any conditional average of the form in \eqref{eq:general_bvp}. Furthermore, estimates of the stationary distribution from an adjoint equation of ${\cal T}^\tau$ can be incorporated into the sampling to accelerate its convergence \cite{copperman2020accelerated,strahan2024bad}.

Given that the committor reports directly on transition progress, it is natural to use its estimates to guide sampling.  
Parrinello and co-workers use an iterative procedure that alternates between training NNs with the Dirichlet form in \eqref{eq:DF_loss}
\cite{kang2024computing,trizio2025everything,kang2026committors,
deng2026role}, or a surrogate for it
\cite{trizio2026ceci,rossi2026let}, and sampling with the bias potential
$-(1/\beta)\ln(|\nabla q_\theta(\bx)|^2+\varepsilon)$, where $\beta$ is the inverse temperature and $\varepsilon$ is a regularization parameter. Related adaptive schemes based on committor-dependent potentials were developed by Lin and Ren \cite{lin2025deep}.
Other approaches improve coverage of the transition region using active
importance sampling with umbrella sampling \cite{rotskoff2022active}, the finite-temperature string method \cite{hasyim2022supervised}, or a generative model \cite{wang2026adaptive}.
Chipot, Roux, and co-workers use the committor estimated with VCN to refine path collective variables \cite{chen2026following,megias2025iterative,talmazan2026from}, building on the committor-consistent string method \cite{he2022committor}.

As previously discussed, both the Dirichlet and its finite-$\tau$ analog, the (E)VCN loss, require reweighting to the stationary distribution, which may offset the advantages of adaptive sampling.
By contrast, estimators that allow for an arbitrary sample distribution allow concentrating initial points in the transition state region ($q\approx 0.5$) or distributing them uniformly over $q$ values. This strategy was pursued in the context of DGA \cite{guo2024dynamics}, the two-trajectory trick \cite{strahan2023predicting}, and fixed-point iteration \cite{mitchell2024committor} (as well as models based on full trajectories \cite{ma2006dynamic,lazzeri2025optimal,breebaart2026understanding}). 
Whether the stability and convenience of the Dirichlet form and (E)VCN losses outweigh sampling flexibility likely depends on system properties, such as the presence of multiple pathways and intermediates, and it would be worth quantitatively characterizing the convergence of competing adaptive sampling procedures for reversible transitions.

\section{Alternative approaches}

\subsection{Shooting-based methods}
\label{sec:shooting}

While this review argues in favor of self-supervised methods for learning the committor owing to the control they afford in sampling, supervised shooting-based methods are used extensively and are under active development \cite{ma2025reaction}.
In particular, the Artificial Intelligence for Molecular Mechanism Discovery (AIMMD) method \cite{jung2023machine} couples a neural network that learns the committor on the fly from the outcomes of two-way shooting moves \cite{bolhuis2002transition} with an adaptive scheme that uses the current committor model to bias the selection of new shooting points toward the transition region.
The network represents the logit of the committor, $\ln[q_\theta(\bx)/(1-q_\theta(\bx))]$, and is trained by maximizing the likelihood of the shooting outcomes treated as Bernoulli trials, which is equivalent to minimizing the binary cross-entropy.
The network input is generally a library of candidate physical variables, from which the most important are identified and used for symbolic regression \cite{jung2023machine}, although graph neural networks, discussed in Section \ref{sec:gnn}, were recently used to remove the need to choose such variables \cite{lichtinger2026accelerated}.
Recent variants combine committor learning with transition interface sampling \cite{breebaart2026understanding,breebaart2026training}, rejection-free shooting-point sampling that is uniform in the logit committor \cite{lazzeri2025optimal}, or generative modeling \cite{negedly2026correlation}. Approximate free energies and rates can also be recovered from the data by reweighting \cite{lazzeri2023molecular,lazzeri2025optimal}.

Methods that regress the committor on shooting outcomes (including but not limited to AIMMD) have been applied to a wide range of systems, including molecular isomerization \cite{ma2005automatic,mori2026deep,topfer2026machine}, polymer and protein folding \cite{jung2023machine,lazzeri2025optimal}, ion association and dissociation \cite{mullen2014transmission,joswiak2018ion,jung2023machine,okada2026deep,mori2026deep}, nucleation \cite{beckham2007surface,beckham2008evidence,lupi2016pre,lupi2017role,jung2023machine}, ligand (un)binding \cite{breebaart2026understanding,breebaart2026training,lichtinger2026accelerated}, molecular recognition for DNA repair \cite{hu2008two}, and membrane-associated processes \cite{jung2023machine,horvath2025stim1,jackel2025free,post2026ai}.
It would be useful to compare supervised methods, trained on shooting outcomes, with self-supervised methods, trained on trajectory segments, in terms of the simulation cost needed to reach a given accuracy in the committor and the rate.
Such comparisons should span transitions with low and high barriers, with single and multiple reaction channels, and with and without intermediates.

\subsection{Dynamics emulators}

An alternative to estimating kinetic statistics directly is to learn a model of the dynamics. Although doing so remains challenging with traditional approaches \cite{lee2019multi,aristoff2023arbitrarily,lin2023regression,dalton2025memory,girardier2026correcting}, machine-learned dynamics emulators are now successfully used for short-term weather prediction \cite{lam2023learning,bi2023accurate}, and they are improving in the molecular sciences \cite{sidky2020latent,vlachas2022accelerated,klein2023timewarp,schreiner2024implicit,hsu2024score,bigi2025flashmd,thiemann2026forcefree,vigueradiez2026transferable}. Unlike machine-learned force fields, which approximate the forces used in numerical integration, emulators attempt to approximate evolution over a time interval $\tau$ much longer than a single MD time step.  Trained on the same short-trajectory data used by the methods in Section~\ref{sec:trajectory}, emulators learn parameterizations of transition probability densities corresponding to the transition operator $\mathcal{T}^\tau$. They can then be applied recursively to generate trajectories longer than $\tau$.  Related generative models learn distributions of entire trajectory segments conditioned on selected frames \cite{jing2024generative}. 

Because they aim to replicate, at lower cost and often for a reduced set of variables, a trajectory that is statistically the same as would be generated by MD simulations, emulators can become trapped in long-lived intermediates, and the trajectories required to estimate the committor and other kinetic statistics by direct shooting can be extremely long. Yet, they often do not maintain statistical accuracy over time scales required to sample rare events \cite{hsu2024score,bigi2025flashmd}. Despite substantial investment, these issues persist even in the weather and climate setting \cite{Baxter2026,Henn2026AIMIP}. Addressing them will require good sampling near transition regions \cite{hsu2024score} (Sections~\ref{sec:sampling} and \ref{sec:generative}) and treatment of memory effects (Section~\ref{sec:nonmarkov}), for the reasons highlighted previously in this review.

\subsection{Committor and optimal control}
\label{sec:control}

Estimates of the committor allow a special form of biased sampling in which every trajectory becomes reactive. Doob's $h$-transform \cite{doob1957conditional} with $q$ as the harmonic function $h$ (see \eqref{eq:Lq=0}) defines the bias drift  
\begin{equation}
  {\bf f}^* = 2D(\bx)\nabla\ln q(\bx) = \frac{2D(\bx)\nabla q(\bx)}{q(\bx)}.
  \label{eq:optimal_control}
\end{equation}
Adding \eqref{eq:optimal_control} to the drift of the overdamped Langevin equation (e.g., $-\beta D(\bx)\nabla V(\bx)$ for a system governed by potential $V(\bx)$) reproduces, exactly and without reweighting, the \emph{reactive} trajectory distribution  of the unbiased dynamics \cite{lu2015reactive}.
The bias drift is proportional to $D(\bx)\nabla q(\bx)$, divided by $q(\bx)$ so that the correction grows large where $q(\bx)$ is small and the transition is least likely. 
Given an accurate approximation of the committor, one can generate trajectories to compute other statistics such as the rate for transition from $A$ to $B$ \cite{yuan2024optimal,singh2024splitting}.
% Cameron and co-workers use this construction directly \cite{yuan2024optimal}.
% Given an accurate approximation of the committor, one can generate trajectories to compute other statistics such as the rate constant for transition from $A$ to $B$; Cameron and co-workers  \cite{yuan2024optimal} and Singh and Limmer \cite{singh2024splitting} do exactly this, solving for the committor with the Dirichlet form. 
Beyond the need for an accurate committor estimate, a practical limitation of this approach is that the reactive trajectories can be long when the pathway involves intermediate basins. In such cases, errors can accumulate and shift the distribution of trajectories and, in turn, estimates of statistics \cite{earle2026relative}.

Among all possible drifts that reliably drive the system from $A$ to $B$,
\eqref{eq:optimal_control} is the one that minimizes the
Bou\'e--Dupuis variational expression (cost functional)
\cite{boue1998variational,chetrite2015variational}
%\begin{equation}\label{eq:BD}
%    -\ln q(\bx) = \min_{\bf f}
%    \mathbf{E}_{\X^{\bf f}_0=x}\left[\frac{1}{2}\int_0^{\T} \lVert {\bf f}(t,\X^{\bf f}_t)\rVert^2 dt\right],
%\end{equation}
%$$
%-\ln q(\bx) = \min_{{\bf f}:\,\X_{\T}^{\bf f}\in B}
%\mathbf{E}_{\X_0^{\bf f}=x}\!\left[\frac{1}{4}\int_0^{\T} {\bf f}(t,\X_t^{\bf f})^\top D(\X_t^{\bf f})^{-1}\,{\bf f}(t,\X_t^{\bf f})\,dt\right],
%$$
\begin{equation}\label{eq:BD}
%-\ln q(\bx) = \min_{{\bf f}}
%\E_{\X_0^{\bf f}=\bx}\!\left[\frac{1}{4}\int_0^{\T} {\bf f}^\top D^{-1}\,{\bf f}\,dt\right],
-\ln q(\bx) = \min_{{\bf f}}
\E\!\left[\frac{1}{4}\int_0^{\T} {\bf f}^\top D^{-1}\,{\bf f}\,dt \,\Big|\, \X_0^{\bf f}=\bx\right],
\end{equation}
where we suppress the arguments of ${\bf f}(t,\X_t^{\bf f})$ and $D(\X_t^{\bf f})$ for clarity, $\X^{\bf f}_t$ is the biased process, and, as
previously, $\T$ is the time of first reaching $A\cup B$. Trajectories with
$\X^{\bf f}_{\T}\notin B$ are assigned infinite cost, so the minimization is effectively over vector functions ${\bf f}(t,\bx)$ that cause the process to end in $B$.
Using Girsanov's theorem and assuming the diffusion tensor is unchanged, one can show that the expectation on the right-hand side of \eqref{eq:BD} is the Kullback--Leibler (KL) divergence of the trajectory distribution of $\X^{\bf f}$ with respect to the trajectory distribution of $\X$. 

% Expression \eqref{eq:BD} offers an alternative route to the committor:  one can optimize a trial drift directly.  Because of the chain rule of KL divergence, the excess cost of a trial drift ${\bf f}$ relative to the optimal drift ${\bf f}^*$ is the KL divergence between the trajectory distributions they generate.
% More generally, the cost of any driven dynamics relative to the unbiased dynamics is a dissipation in units of thermal energy, related to a bound on rate enhancement established by Kuznets-Speck and Limmer \cite{kuznetsspeck2021dissipation}.  
% Several approaches exploit \eqref{eq:BD} to estimate a committor and obtain the Doob $h$-transform from its gradient \cite{hartmann2019variational,singh2023variational}; others fit the control drift or trajectory distribution directly against a loss evaluated on full transition paths \cite{hartmann2012efficient,das2019variational,holdijk2023stochastic,du2024doob}. These approaches are closely related to importance sampling, Schr\"odinger bridges, and variational path sampling and are reviewed in Ref.~\citenum{singh2025variational}. Du and co-workers instead optimize the control via a self-consistency relation analogous to the fixed-point iteration of Section~\ref{sec:richardson} \cite{du2026rare}.

Expression \eqref{eq:BD} offers an alternative route to the committor:  one can optimize a trial drift directly.  Because of the chain rule of KL divergence, the excess cost of a trial drift ${\bf f}$ relative to the optimal drift ${\bf f}^*$ is the KL divergence between the trajectory distributions they generate.
More generally, the cost of any driven dynamics relative to the unbiased dynamics is a dissipation in units of thermal energy, related to a bound on rate enhancement established by Kuznets-Speck and Limmer \cite{kuznetsspeck2021dissipation}.  
Hartmann and co-workers \cite{hartmann2019variational} and Du and co-workers \cite{du2026rare} use the hitting-time control problem in \eqref{eq:BD} to estimate the committor or its associated optimal control; the latter use a self-consistency relation analogous to the fixed-point iteration of Section~\ref{sec:richardson}. Related finite-time control approaches condition trajectories on being inside $B$ at a prescribed terminal time \cite{holdijk2023stochastic,du2024doob,singh2023variational,singh2024splitting}.
Others optimize control forces directly for rare-event and path sampling
\cite{hartmann2012efficient,das2019variational}. These approaches are closely related to Schr\"odinger bridges, importance sampling, and variational path sampling, the latter two of which are reviewed in Ref.~\citenum{singh2025variational}.

\section{Outlook}

Over the past two decades, studies that compute the committor and other mechanistically informative kinetic statistics have progressed from estimates with large uncertainties for tens of microscopic states to estimates with small uncertainties for millions of states; this transition enables qualitatively new types of analyses as discussed in Section \ref{sec:applications}. This progress reflects the theoretical and computational advances described above, together with continued growth in computational power. An important remaining task is for researchers to deploy and evaluate these methods on the diverse chemical systems that motivate their research. We noted a number of specific open problems along the way; here, we highlight a few additional areas that we expect to be particularly fruitful in the near term.

\subsection{Connecting to reinforcement learning}

The stochastic optimal control methods discussed in Section \ref{sec:control} seek to learn bias forces that make all trajectories reactive. This goal is shared by reinforcement learning (RL), which seeks to learn policies that map states to actions in Markov decision processes so as to maximize expected cumulative rewards.  The RL connection to the committor more generally can be made precise:  the committor is the expected reward (value function) when the reward is 1 on $B$ and 0 otherwise (with no discounting).  In the language of RL, DGA is temporal difference (TD) learning with linear function approximation \cite{bradtke1996linear,boyan2002technical}.  The two-trajectory trick is a form of double sampling, which was introduced in RL to address an analog of the gradient estimation  difficulty discussed in Section~\ref{sec:FP} \cite{baird1995residual}.

%[Version 1] The RL connection points to potential problems for computing the committor and other kinetic statistics.  The ``deadly triad'' of function approximation,  bootstrapping (updating an estimate of a statistic using another estimate of that same statistic, which arises here in the fixed-point iteration in Section \ref{sec:richardson}), and off-policy data (trajectory data sampled under arbitrary $\mu(\bx)$) is well known to lead to numerical instability \citep{sutton2018reinforcement}. By the same token, RL may also point toward solutions, given the many algorithms that successfully learn policies despite the deadly triad, as well as ways to derive efficiency guarantees like those in Section~\ref{sec:guarantees}. We believe that elaborating the RL connection can lead to robust sampling and estimation strategies; methods developed for molecular dynamics can also inform RL, e.g., by improving learning when rare events are important.

%[Version 2] 
The RL connection also raises a potential tension between statistical efficiency and numerical stability.  As discussed in Section~\ref{sec:guarantees}, efficient estimation of transition statistics can require sampling initial conditions from a distribution $\mu(\bx)$ that differs substantially from the stationary distribution $\pi(\bx)$.  The fixed-point iteration in Section~\ref{sec:richardson} combines this freedom with function approximation and bootstrapping, since the current estimate $q_{\theta_k}(\bx)$ is used to construct the target for $q_{\theta_{k+1}}(\bx)$.  This setting is closely related to the ``deadly triad'' in RL.  When fixed-point iterations are projected, mismatch between the sampling distribution and the dynamics can invalidate standard contraction arguments for convergence guarantees, and worst-case examples can diverge \cite{sutton2018reinforcement,tsitsiklis1997analysis}.  This does not imply that the fixed-point scheme is unstable in practice but suggests that sampling strategies chosen for statistical efficiency should also be examined for their effect on numerical stability.  RL may provide useful tools for understanding and controlling this tradeoff.

\subsection{Reducing feature engineering}
\label{sec:gnn}
%Existing NNs for the committor are largely multilayer perceptrons that take as inputs either Cartesian coordinates for aligned structures or internal coordinates. However, each has its limitations:  how best to align structures is often unclear, and small differences in internal coordinates can correspond to large differences in molecular structures. 
Existing NNs for the committor rely on manually selected features or on Cartesian coordinates aligned on manually selected atoms. Equivariant geometric graph NNs (GNNs), in which information is passed between graph nodes that correspond to atom positions, are promising alternatives because they can take Cartesian coordinates as inputs while respecting molecular symmetries. In principle, then, the analysis can proceed from atomic coordinates alone, and the features relevant to the dynamics can be learned rather than chosen in advance. GNNs were used to compute committors for a variety of peptide conformational transitions (including Trp-cage folding), chemical reactions (Diels--Alder), and associations/dissociations (NaCl, CaCO$_3$, and host-guest and protein-ligand systems) \cite{kang2026committors,contrerasarredondo2026learning,lichtinger2026accelerated}. GNNs were also used to learn state definitions \cite{zou2025graph,pengmei2025using,pengmei2026hierarchical}, which can strongly affect estimates of kinetic statistics.

However, GNNs can be memory- and runtime-limited. This cost is compounded when avoiding data leakage requires splitting training and validation sets by trajectory rather than by time-lagged pair, because each trajectory-level split must then use a large minibatch to preserve sampling diversity \cite{pengmei2025using,pengmei2026hierarchical}. Two developments can extend the tractable number of nodes by orders of magnitude. One pre-trains a GNN on a tractable task and then fixes its parameters, optimizing only an output head specific to the dynamics analysis \cite{pengmei2025using}; the other merges information across windows of residues without compromising model expressiveness \cite{pengmei2026hierarchical}. How best to ensure stability when training such expressive models is an open question, as is how to interpret them. 

\subsection{Interpretation}

Presently the main route to interpreting the committor is to choose a library of candidate physical variables and fit a model that depends on a small number of those variables to either shooting data \cite{ma2005automatic,jung2023machine,post2026ai,breebaart2026training,topfer2026machine,ma2025reaction} or a model of the committor obtained from the methods described in Sections \ref{sec:PDE} and \ref{sec:trajectory} \cite{guo2024dynamics}. Correlations with the committor \cite{jeong2024analysis}, gradient-based sensitivity analyses \cite{breebaart2026understanding}, and systematic approaches for identifying important features  \cite{mori2026deep} such as Local Interpretable Model-agnostic Explanations (LIME) \cite{ribeiro2016should} and SHapley Additive exPlanations (SHAP) \cite{lundberg2017unified} are also used. Though reasonable, these approaches can only discover variables in the library, and human intuition is still needed to translate the  identified variables into mechanisms. Methods that allow one to discover new features are an active area of research. Sparse autoencoders were recently used to learn features from trained models for protein structure and binding prediction \cite{simon2025interplm,lu2025mapping}, and it would be interesting to apply them in the context of dynamics. 
The average gradient outer product (AGOP), used to explain how neural networks learn features \cite{radhakrishnan2024mechanism}, was recently applied to extract physically interpretable feature directions directly from a neural network trained to forecast ocean-atmosphere dynamics \cite{hui2026explaining}.
Closely related decompositions of rates and entropy production into contributions from specified coordinates already exist for stochastic dynamics \cite{singh2023variational,boffi2024deep}; for Hamiltonian dynamics, Ma and co-workers decompose potential and kinetic energy flows and obtain coordinates from singular vectors of a generalization of these flows, the generalized work functional \cite{li2016reaction,li2020kinetic,wu2022rigorous,wu2022exact,ma2025reaction}. Using the eigenvectors of the AGOP for a model of the committor could offer a data-driven alternative to a predefined variable library, although the resulting directions would still require interpretation.
%Applying the gradient outer product to a model of the committor could offer a data-driven alternative to a predefined variable library, although the result would still require interpretation.
Given advances in agentic hypothesis generation \cite{villaescusa2025denario,gridach2025agentic,gottweis2026accelerating}, we expect automated reasoning about candidate mechanisms to soon complement feature-importance and feature-discovery methods.
% Lu et al. is placeholder for paper we are revising now with SAE

\subsection{Generative machine learning}
\label{sec:generative}
The methods discussed in Sections \ref{sec:PDE} and \ref{sec:trajectory} regress on data to estimate kinetic statistics. Generative machine learning methods instead learn distributions. They can be used to sample initial points \cite{noe2019boltzmann,falkner2023conditioning,tang2026breaking,negedly2026correlation,pengmei2026recursive} and, because successive points sampled are uncorrelated, they complement the MD-based approaches discussed in Section \ref{sec:samplingpractice}, which must traverse distributions. However, nothing guarantees that a generative model's learned distribution corresponds to a physically meaningful statistical ensemble, and such models can miss important states entirely \cite{felardos2023designing,wang2026mitigating}. Reweighting the generated ensemble after the fact is ineffective when it overlaps little with the target, which is the same obstacle that makes equilibrium sampling inefficient for committor estimation (Section \ref{sec:sampling}).

%Recent studies address these issues by coupling generative models with committor learning in self-consistent loops to concentrate sampling near $q\approx 0.5$, instead of training on an equilibrium (Boltzmann) ensemble \cite{tang2026breaking,negedly2026correlation}; the Gen-COMPAS method \cite{tang2026breaking}, which combines diffusion models with VCN, is demonstrated on Trp-cage folding, coupled folding and ligand binding of ribose-binding protein, and conformational transitions of three membrane proteins:  the mitochondrial ADP/ATP carrier (AAC), nicotinic acetylcholine receptor (nAChR), and a vacuolar ATPase.
%For AAC, the predicted landscape shows that the transport-competent pathway depends on the bound substrate: removing it eliminates the intermediate state and raises the barrier substantially, linking substrate occupancy directly to the transport mechanism. For nAChR, Gen-COMPAS identifies an asymmetric intermediate and an ordered activation sequence that matches independently reported cryo-EM and electrophysiology data \cite{thompson2025asynchronous}.

Recent studies address these issues by coupling generative models with committor learning in self-consistent loops to concentrate sampling near $q\approx 0.5$, instead of training on an equilibrium (Boltzmann) ensemble \cite{tang2026breaking,negedly2026correlation}; the Gen-COMPAS method \cite{tang2026breaking}, which combines diffusion models with VCN, is demonstrated on Trp-cage folding, coupled folding and ligand binding of ribose-binding protein, and conformational transitions in three membrane proteins (the mitochondrial ADP/ATP carrier, the nicotinic acetylcholine receptor, and a vacuolar ATPase), recovering a substrate-dependent transport mechanism and an asymmetric receptor-activation pathway subsequently corroborated by an independent experiment \cite{thompson2025asynchronous}.

The examples immediately above leverage MD simulations for transition-path sampling, but in principle, generative models can also be used directly 
\cite{lelievre2023generative,jing2024generative,triplett2025diffusion,raja2025action}. 
However, models trained on Boltzmann-weighted configurations are unreliable near transition states because these regions are poorly represented in the training data. Approaches now being developed to address this problem follow two general strategies. The first is that, as espoused throughout this review, one should place data where accuracy is needed: training on restrained distributions concentrates samples in selected regions, and an appropriately conditioned noising process removes the resulting bias, so that the model recovers the unrestrained target without importance weights \cite{zhang2026restrained}. The second is that physical consistency can be enforced during inference rather than learned: alternating generative and MD updates in a Gibbs sampler composes a pretrained model for selected variables with explicit context for the rest, with conditions under which this is exact \cite{wang2026composing}, in contrast to heuristic guidance terms. Because this construction requires only local conditionals, it extends to any path ensemble with Markov structure \cite{wang2026quantum}. Outperforming MD is likely to require combining such strategies, particularly once the cost of generating training data is taken into account, and whether transferable models can offset this cost remains to be shown.

\section{Conclusions}

In this review, we discussed methods for computing kinetic statistics that reveal microscopic mechanisms. A recurring theme was the advantage of estimating these statistics directly from trajectory data rather than from an assumed form for the dynamics; this approach extends naturally to systems for which the form of the dynamics is unknown, including reduced sets of variables, irreversible dynamics, and experimental time series. Key technical points were the importance of stopping trajectories at the boundaries of the metastable states and achieving good sampling in transition regions. 
We discussed how the committor can be used not only to analyze existing data but also to guide sampling, and we expect fully realizing the latter will require systematically comparing existing approaches and developing new ones.

We focused on molecular examples, but the underlying framework is general. Indeed, the methods we discussed are already used to study transitions in atmosphere and ocean data \cite{finkel2021learning,finkel2023revealing,finkel2023data,jacquesdumas2023data,miron2021transition,drouin2022surface,miron2022nadw}. Beyond reinforcement learning, developments in machine learning more broadly, from equivariant graph architectures to generative models, are already shaping analyses of dynamics, and continued exchange in both directions can sharpen sampling strategies, model design, and the theoretical guarantees that accompany them. Much of the progress in computing kinetic statistics made to date reflects a dialog among chemists, physicists, statisticians, applied mathematicians, and computer scientists. We hope this review lowers the barriers that separate these communities and helps researchers apply the methods described here to diverse new systems.

\section*{Acknowledgments}

We thank the many members of the Weare and Dinner groups who contributed to work described in this review and discussions related to it. We thank Qiang Cui for suggesting this review, Peter Bolhuis, Chris Chipot, Roberto Covino, Yuanqi Du, Chatipat Lorpaiboon, Simon Olsson, and Aditya Singh for comments on the manuscript, and Caitlin McCandler for help with Fig.\ \ref{fig:catalysis}. We acknowledge support from National Institutes of Health awards R35 GM136381 and R01 GM169399, National Science Foundation award DMS-2054306, and the Margot and Tom Pritzker Science Foundation.

%\bibliographystyle{unsrt}
%\bibliography{master2}

\end{document}